\documentclass[3p, preprint]{elsarticle}
\usepackage[dvipsnames]{xcolor}
\usepackage[utf8]{inputenc}
\usepackage[T1]{fontenc}
\usepackage{bm}
\usepackage{booktabs}
\usepackage{tcolorbox}
\usepackage{longtable}
\usepackage{multirow}
\usepackage{pdflscape}
\usepackage{amssymb}
\usepackage[normalem]{ulem}
\usepackage{amsmath}
\usepackage{lscape}
\usepackage{url}
\usepackage{enumitem}
\usepackage{gensymb}
\usepackage{hyperref}
\usepackage[misc]{ifsym}
\usepackage{annotate-equations}
\definecolor{cadmiumgreen}{rgb}{0.0, 0.42, 0.24}

\newtcolorbox{concept}[1]{
    colback=blue!10!white,
    colframe=blue!50!black,
    title={\textbf{Paragraph}: #1},
}

\newtcolorbox{concept2}[1]{
    colback=green!10!white,
    colframe=green!50!black,
    title={\textbf{Paragraph}: #1},
}

\newtcolorbox{important}[1]{
    colback=red!10!white,
    colframe=red!50!black,
    title={\textbf{Important}: #1},
}

\journal{arXiv}

\begin{document}

\begin{frontmatter}
% artIfacts not artEfacts
\title{Artifacts in Photoacoustic Imaging: Origins and Mitigations}

\author[aff1,equal]{Max T Rietberg}
\author[aff2,aff3,aff5,equal]{Janek Gr\"ohl}
\author[aff2,aff3,equal]{Thomas R Else}
\author[aff2,aff3]{Sarah E Bohndiek}
\author[aff1]{Srirang Manohar\corref{cor1}}
\author[aff4]{Benjamin T Cox\corref{cor1}}
\affiliation[aff1]{organization={Multi-Modality Medical Imaging, TechMed Centre, University of Twente},
            addressline={Drienerlolaan 5},
            city={Enschede},
            postcode={7522 NB},
            state={Overijssel},
            country={The Netherlands}}
\affiliation[aff2]{organization={Cancer Research UK, Cambridge Institute, University of Cambridge},
            addressline={Robinson Way},
            city={Cambridge},
            postcode={CB2 0RE},
            country={United Kingdom}}
\affiliation[aff3]{organization={Department of Physics, University of Cambridge},
            addressline={JJ Thomson Avenue},
            city={Cambridge},
            postcode={CB3 0HE},
            country={United Kingdom}}
\affiliation[aff4]{organization={Department of Medical Physics and Biomedical Engineering, University College London},
            addressline={Gower St},
            city={London},
            postcode={WC1E 6BT},
            country={United Kingdom}}
\affiliation[aff5]{organization={ENI-G, a Joint Initiative of the University Medical Center Göttingen and the Max Planck Institute for Multidisciplinary Sciences},
            addressline={Grisebachstr. 5},
            city={Göttingen},
            postcode={37077},
            country={Germany}}

\fntext[equal]{Equal contribution}
\cortext[cor1]{Corresponding Authors: SM (s.manohar@utwente.nl) and BTC (b.cox@ucl.ac.uk)}

\begin{abstract} %150 words maximum
Photoacoustic imaging (PAI) is rapidly moving from the laboratory to the clinic, increasing the need to understand confounders which might adversely affect patient care. Over the past five years, landmark studies have shown the clinical utility of PAI, leading to regulatory approval of several devices. In this article, we describe the various causes of artifacts in PAI, providing schematic overviews and practical examples, simulated as well as experimental. This work serves two purposes: (1) educating clinical users to identify artifacts, understand their causes, and assess whether their impact, and (2) providing a reference of the limitations of current systems for those working to improve them. We explain how two aspects of PAI systems lead to artifacts: their inability to measure complete data sets, and embedded assumptions during reconstruction. We describe the physics underlying PAI, and propose a classification of the artifacts. The paper concludes by discussing possible advanced mitigation strategies.
\end{abstract}

%% Keywords

\end{frontmatter}

% \tableofcontents

\section{Introduction}
Medical imaging technologies enable visualization of the structure and function of biological tissues, thereby facilitating disease detection, diagnosis, monitoring of treatment response and follow-up. Medical imaging modalities rely on the measurement of transmitted, absorbed, scattered, or emitted energy of some form. Measured data and physical models of energy-tissue interactions are then used to reconstruct images~\cite{suetens2017fundamentals}. These models necessarily involve simplifying assumptions, and so describe the true physical processes only approximately. Furthermore, during data collection some information will be lost, whether due to hardware limitations or subject and environmental factors. The reconstructed image will therefore not be a perfect representation of the underlying tissue but will contain deviations, called artifacts. An artifact is a structure, distortion, or feature in an image that does not correspond to the actual anatomical, morphological, physiological, or pathological reality of the imaged subject~\cite{sureshbabu2005pet,bellon1986mr}. The presence of artifacts may confound clinical interpretation and negatively affect decision making. For example, artifacts includes the obscuring, displacement or distortion of genuine structural or functional features, or the presence of features that are not real, perhaps giving the illusion of pathophysiology. Artifacts can lead to images being misinterpreted, errors in quantitative (e.g.\ molecular) information extracted from the image,  and false positive or false negative imaging biomarkers. A thorough understanding of artifacts in medical imaging is therefore crucial to ensure reliable and accurate clinical use.

Artifacts can be especially prominent in multi-physics imaging modalities like photoacoustic imaging. In photoacoustic (PA) imaging (PAI) the acoustic emissions from the absorbed optical energy distribution are measured, in contrast to purely optical imaging methods such as diffuse optical tomography~\cite{boas2001imaging}, where transmitted or reflected light is detected. In PAI, tissue is interrogated with nanosecond pulsed light, which is scattered and absorbed inside the tissue. The light is absorbed by chromophores, leading to a localized temperature and pressure rise, and subsequently, due to thermoelastic expansion, the emergence of acoustic pressure waves~\cite{wang2007biomedical}. Detection of acoustic waves outside the tissue followed by image reconstruction enables visualization of the distribution of absorbed optical energy up to several centimeters deep. By varying the wavelength of incident light, absorption of several different molecules can be probed~\cite{wang2012photoacoustic}. Typical features extracted from PAI with promise for clinical translation include (1) vascular architecture~\cite{chen2021dedicated,brown2022quantification,attia2019review}, (2) signal intensity at a target wavelength~\cite{jo2017functional,knieling2017multispectral,regensburger2021optoacoustic}, (3) blood oxygen saturation~\cite{li2018photoacoustic,grohl2021learned,vogt2019photoacoustic}, (4) general molecular concentration information of, e.g., fatty tissue, glandular tissue, water~\cite{ntziachristos2010molecular,yao2018recent}, and (5) contrast agent distribution~\cite{weber2016contrast,fu2019photoacoustic}. Photoacoustic image formation thus involves an interplay of optical, thermal, elastic and acoustic properties, and gives rise to a variety of unique artifacts not seen in optical or ultrasound imaging separately. The presence of artifacts can have negative effects on the qualitative and quantitative interpretation of PAI images and their extracted features.
\begin{center}
\begin{tcolorbox}[width=0.8\linewidth]
\centering
\textbf{In the context of PAI, we define an artifact as a feature in the image that does not correspond, spatially or spectrally, to the true anatomical, morphological and spectral characteristics of the imaging target under investigation.}
\end{tcolorbox}
\end{center}

As PAI is on the cusp of achieving clinical breakthroughs in various applications~\cite{lin2022emerging, karlas2021optoacoustic}, we believe it would be helpful to raise awareness of artifacts in the PAI community, especially among the growing number of clinical PAI users. PA users should be able to identify artifacts, so they can better understand and improve the interpretation of \textit{in vivo} PA images or even take advantage of the artifacts, as they could carry diagnostic information. Furthermore, we hope that recognizing artifacts and understanding their origins will help guide the development of PAI systems towards mitigations and hasten the progress towards improved PAI technology.

In this work, we structure the manuscript by the \textit{sources} and \textit{causes} of artifacts. Starting with a short description of the physics involved in PAI, we continue by outlining the reasons behind the emergence of each artifact, and briefly discuss possible mitigation strategies. Each artifact is demonstrated with simulated images and, where available, experimental examples. Simulations are performed on simple digital phantoms in which the assumptions made during image reconstruction are deliberately broken to demonstrate their confounding effect. For illustrative purposes, the simulations are restricted to simple but typical clinical PA detection geometries; however, similar artifacts would appear in other hardware configurations~\cite{oraevsky2018clinical, noltes2023towards}. We conclude with a discussion of possible advanced artifact mitigation strategies. 

\section{Artifact Classification}
Necessary compromises in the design of PAI systems mean that most PA images contain artifacts. Taking a high-level view, there are two parts to any PAI system: hardware for data acquisition, and a method for reconstructing and displaying the images. There are two corresponding classes of underlying causes of artifacts: (1) insufficient data, e.g.\ limited array size or limited bandwidth, and (2) incorrect assumptions in the image reconstruction algorithm, e.g.\ inaccurate approximations to the physics of wave propagation or an overly simple model of the detector response. 

In practical applications, there are often good reasons for designing an imaging system that measures incomplete data, or for developing an image reconstruction approach that makes inaccurate assumptions. First, there are \textit{clinical requirements}. These are dictated by what the clinician or researcher needs to see, which requires the instrument to access the body part of interest with a sufficient field-of-view and resolution. Further challenges, such as restrictions on the clinical workspace, fitting around other devices, and patient safety and comfort, are also key considerations. Second, there are \textit{physics constraints}, e.g.\ limited light penetration or acoustic attenuation that limit imaging depths. Third, there are \textit{practical limitations}, which include the availability of resources (raw materials, fabrication facilities, computing hardware), financial costs of hardware and operation, or even intellectual property rights and market demand. Trade-offs must be made in imaging system design, leading to widely differing PAI systems from microscopes to large-area bowls for breast imaging. 

\begin{figure}
    \centering    \includegraphics[width=.7\linewidth]{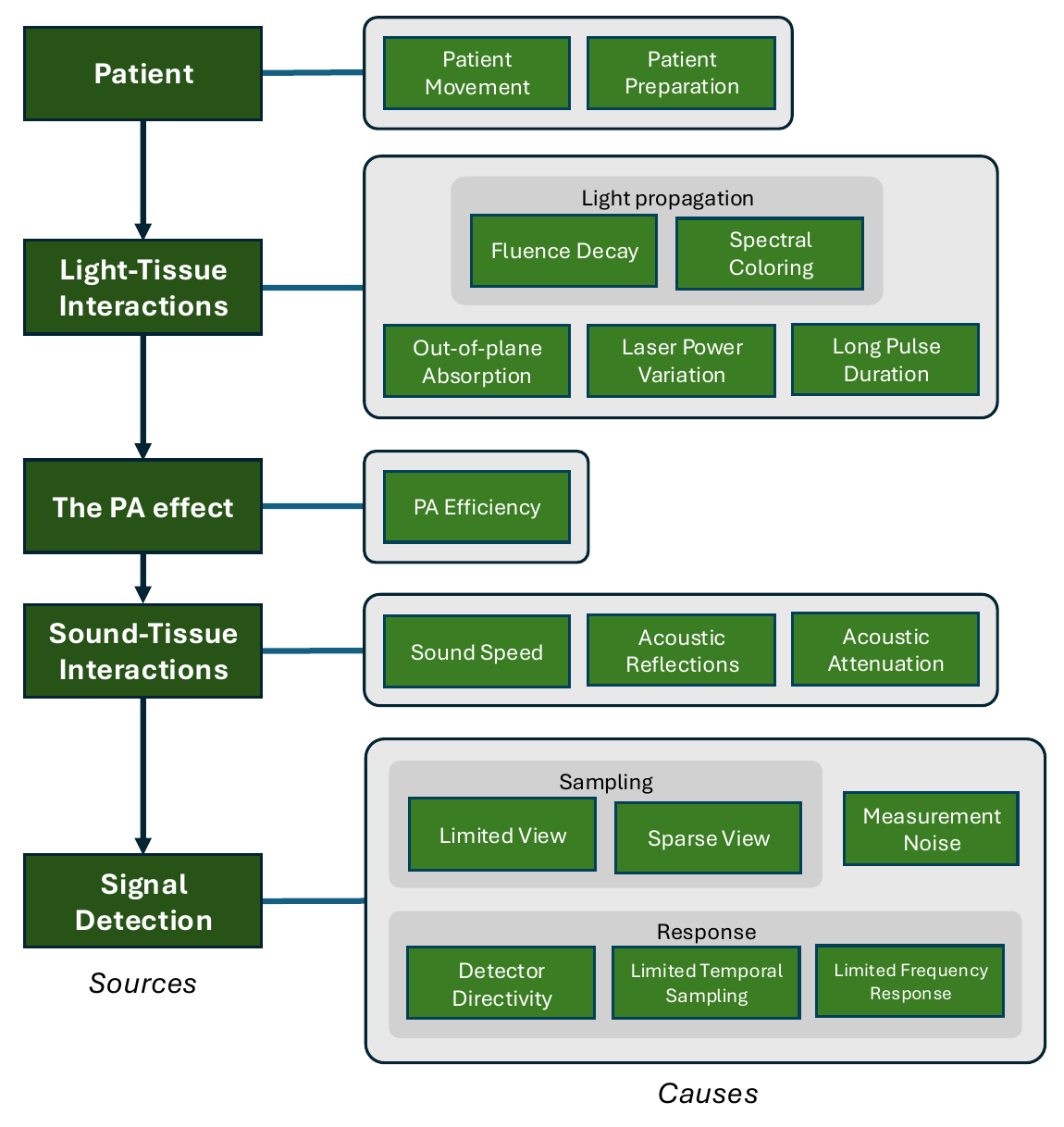}
    \caption{\textbf{Overview of the artifacts treated in this paper.} They are arranged by the point at which they occur throughout the photoacoustic (PA) signal generation process: induced by (1) the patient or by the preparation of the patient; (2) the interactions of light with tissue; (3) the conversion of heat into sound energy; (4) the interactions of sound with tissue; (5) the detection of the sound waves.}
    \label{fig:overview}
\end{figure}

Artifacts in PAI may be classified in many different ways. Our approach is to classify artifacts into five groups distinguished by the \textit{source} of the artifact, namely:
\begin{enumerate}
    \item Patient: artifacts resulting from subject movement and experimental conditions.
    \item Light-Matter Interactions: artifacts arising from inaccurate assumptions about the interactions between light and tissue.
    \item The Photoacoustic Effect: artifacts arising from inaccurate assumptions relating to the photoacoustic efficiency (the \textit{Gr\"uneisen parameter}).
    \item Sound-Matter Interactions: artifacts arising from inaccurate assumptions about the interactions between sound and tissue.
    \item Signal Detection: artifacts arising from shortcomings in acquiring data without loss or distortion of information.
\end{enumerate}
These artifact \textit{sources} can be subdivided into what we call artifact \textit{causes}, such as fluence decay or acoustic reflection. An overview is given in Fig.~\ref{fig:overview}. 
In addition, artifacts can be sorted into different types categorized by their appearance in the image. 
One way to classify the \textit{effects} of artifacts is: (1) dislocation: a shift or displacement of image features from their correct positions, (2) splitting: duplication or division of a single object or feature, (3) blurring of image features: loss of sharpness or detail in the image, (4) the emergence of clutter superimposed on the image: unwanted visual elements that appear over the image, and (5) the unexpected loss or systematic change of the signal: portions of the image data go missing or change consistently in a way not intended. See Tab.~\ref{tab:assumptions} and \ref{tab:artifact_list} in the Supplementary information for the relationships between assumptions, the artifact \textit{sources} and \textit{causes}, and the \textit{effects}.

\clearpage % force Latex to drop any held images here, so we can keep the sections separate while editing.

\section{PAI Imaging Physics}
The following section describes the underlying physical principles which may be of interest to those interested in a deeper understanding. However, it can be skipped by readers who are primarily focused on a practical understanding of PA artifacts.

\subsection{Light Propagation and Absorption}
The behavior of light in scattering media such as human tissue can, for the most part, be described accurately by modeling light as a particle~\cite{wang2007biomedical}. Modeling can be achieved using the radiative transfer equation, also called Boltzmann's transport equation for low-energy, monochromatic photons. The radiative transfer equation describes the light radiance, $\phi(\mathbf{x}, \hat{\mathbf{s}}, t)$, the rate of optical energy flow per unit area per unit solid angle in direction $\hat{\mathbf{s}}$ at position $\mathbf{x}$ at time $t$, and can be written~\cite{sandell2011review} as:
\vspace{3em}
\begin{align}
\begin{split}
\frac{1}{c} \eqnmarkbox[Cerulean]{lhs}{\frac{\partial \phi(\mathbf{x}, \hat{\mathbf{s}}, t)}{\partial t}}
=
\eqnmarkbox[WildStrawberry]{source}{q(\mathbf{x}, \hat{\mathbf{s}}, t)}
&-
\eqnmarkbox[Plum]{space}{\hat{\mathbf{s}} \cdot \nabla \phi(\mathbf{x}, \hat{\mathbf{s}}, t)}
-
\eqnmarkbox[RedOrange]{away}{\left(\mu_a(\mathbf{x})+\mu_s(\mathbf{x})\right)\phi(\mathbf{x}, \hat{\mathbf{s}}, t)}\\
&+
\eqnmarkbox[Green]{into}{\mu_s \int \Theta\left(\hat{\mathbf{s}}, \hat{\mathbf{s}}^{\prime}\right) \phi\left(\mathbf{x}, \hat{\mathbf{s}}^{\prime}, t\right) \mathrm{d} \hat{\mathbf{s}}^{\prime},}
\label{eq:RTE_time}
\end{split}
\end{align}
\annotate[yshift=2em]{above}{lhs}{Change in radiance over time}
\annotate[yshift=1em]{above}{source}{Light source}
\annotate[yshift=1em]{above}{space}{Change in \\ \sffamily\footnotesize radiance over space}
\annotate[yshift=1em]{above}{away}{Absorption and scattering \\ \sffamily\footnotesize away from direction $\hat{\mathbf{s}}$}
\annotate[yshift=0em]{below}{into}{Scattering into direction $\hat{\mathbf{s}}$ from other directions}
\vspace{1em}

\noindent where $c$ is the velocity of light in the medium, $q(\mathbf{x}, \hat{\mathbf{s}}, t)$ is a source of photons, $\mu_a$ is the optical absorption coefficient, $\mu_s$ is the optical scattering coefficient, and $\Theta\left(\hat{\mathbf{s}}, \hat{\mathbf{s}}^{\prime}\right)$ is the scattering phase function (a probability density function describing how photons traveling in direction $\hat{\mathbf{s}}$ will scatter to direction $\hat{\mathbf{s}}^{\prime}$). The radiative transfer equation does not take into account inelastic scattering, radiative losses, wave effects, polarization, ionization, or other photochemical reactions. The absorbed energy density within the tissue $H(\mathbf{x})$, due to the absorption of light, can be written as 

\vspace{2em}
\begin{align}
\begin{split}
    \eqnmarkbox[Cerulean]{lhs}{H(\mathbf{x})} = 
    \eqnmarkbox[WildStrawberry]{mua}{\mu_a(\mathbf{x})}\cdot
    \eqnmarkbox[RedOrange]{phi}{\int\int \phi(\mathbf{x}, \hat{\mathbf{s}}, t) \mathrm{d} \hat{\mathbf{s}}\ dt} 
    = 
    \eqnmarkbox[WildStrawberry]{}{\mu_a(\mathbf{x})}\cdot
    \eqnmarkbox[RedOrange]{}{\Phi(\mathbf{x})},
    \label{eq:absorbed_energy}
\end{split}
\end{align}
\annotate[yshift=2.5em]{above}{lhs}{Absorbed energy density}
\annotate[yshift=1em]{above}{mua}{Optical absorption coefficient}
\annotate[yshift=-0.5em]{below}{phi}{Radiance integrated over direction $\hat{\mathbf{s}}$ and time $t$}
\vspace{1em}

\noindent where the inner integral is over all angles and the time integral is long enough so all photons have either been absorbed or left the tissue. The time and angle-integrated radiance, $\Phi$, is called the fluence.

\subsection{Photoacoustic Effect}
The deposition of optical energy as heat in the tissue gives rise to local increases in temperature and pressure. When the pulse of light, and its subsequent absorption and thermalization, occurs on a timescale $\tau_p$ that is much shorter than the characteristic thermal relaxation time, the condition called \textit{thermal confinement} is satisfied: $\tau_p\ll \tau_h=d_c^2/\alpha$, where $d_c$ is the desired spatial resolution and $\alpha$ is the thermal diffusivity. A more stringent requirement is that the optical pulse is shorter than the acoustic relaxation time, which is called the \textit{stress confinement}, $\tau_p<\tau_s = d_c/v$, where $v$ is the sound speed. Satisfying these conditions means that pressure changes are dissipated as pressure waves and the acoustic propagation can be modeled as an initial value problem, where the initial acoustic pressure distribution, $p_0(\mathbf{x})$, is given by

\clearpage

\vspace{2.5em}
\begin{equation}
    \eqnmarkbox[Green]{p0}{p_0(\mathbf{x})} = 
    \eqnmarkbox[Plum]{gamma}{\Gamma(\mathbf{x})} \cdot
    \eqnmarkbox[Cerulean]{h}{H(\mathbf{x})} = 
    \eqnmarkbox[Plum]{}{\Gamma(\mathbf{x})} \cdot \eqnmarkbox[WildStrawberry]{mua}{\mu_a(\mathbf{x})} \cdot
    \eqnmarkbox[RedOrange]{phi}{\Phi(\mathbf{x})}, 
    \label{eq:initial_pressure}
\end{equation}
\annotate[yshift=2.5em]{above}{gamma}{Gr\"uneisen Parameter}
\annotate[yshift=1em]{above}{h}{Absorbed energy density}
\annotate[yshift=-0.5em]{below}{mua}{Optical absorption coefficient}
\annotate[yshift=1em]{above}{phi}{Light fluence}
\annotate[yshift=-0.5em]{below}{p0}{Initial pressure distribution}
\vspace{1em}

\noindent where $\Gamma$ is the PA pressure efficiency, which is equal to the \textit{Gr\"uneisen parameter} for a pure optically-absorbing fluid. In this case, $\Gamma$ depends on the tissue’s thermomechanical properties, specifically the thermal expansion coefficient $\beta$, isothermal compressibility $K_T$, specific heat capacity $C_p$, and mass density $\rho$, as:

\vspace{3em}
\begin{equation}
\eqnmarkbox[Plum]{Gamma}{\Gamma(\mathbf{x})} = 
\frac{
\eqnmarkbox[Green]{beta}{\beta(\mathbf{x})}
}{
\eqnmarkbox[WildStrawberry]{CP2}{C_p(\mathbf{x})} \cdot \eqnmarkbox[Cerulean]{KT}{K_T(\mathbf{x})} \cdot \eqnmarkbox[RawSienna]{rho}{\rho(\mathbf{x})}} = \frac{\eqnmarkbox[Green]{beta}{\beta(\mathbf{x})}  \cdot \eqnmarkbox[RedOrange]{sos2}{v^2(\mathbf{x})}}{\eqnmarkbox[WildStrawberry]{CP}{C_p(\mathbf{x})}},
\end{equation}
\annotate[yshift=1em]{above}{sos2}{Sound speed}
\annotate[yshift=1.5em]{above}{Gamma}{Gr\"uneisen parameter}
\annotate[yshift=-2em]{below}{KT}{Isothermal compressibility}
\annotate[yshift=-0.5em]{below}{rho}{Density}
\annotate[yshift=-0.5em]{below}{CP}{Specific heat capacity}
\annotate[yshift=2.5em]{above}{beta}{Thermal expansion coefficient}
\vspace{3em}\\
where $K_T =1 / \rho v^2 $, with $v$ as the sound speed.

\subsection{Acoustic Propagation and Detection}
Photoacoustically generated waves are of sufficiently low amplitude that they can be modeled using linear acoustic theory. The PA initial value problem can therefore be written using the wave equation for heterogeneous media:

\vspace{1.5em}
\begin{equation}
    \frac{1}{\eqnmarkbox[RedOrange]{sos}{v^2(x)}} \cdot
    \frac{\partial^2 \eqnmarkbox[Plum]{wave}{p}}{\partial t^2} - 
    \nabla^2\eqnmarkbox[Plum]{wave2}{p} + 
    \eqnmarkbox[Cerulean]{rho}{\frac{\nabla \rho_0 \cdot \nabla p}{\rho_0}} + 
    \eqnmarkbox[WildStrawberry]{atn}{\mathcal{L}p}
     = 0, 
     \label{eq:wave_equation}
\end{equation}
\annotate[yshift=-0.5em]{below}{sos}{Sound speed}
\annotate[yshift=1.5em]{above}{atn}{Acoustic absorption}
\annotate[yshift=-0.5em]{below}{rho}{Influence of density}
\annotate[yshift=1em]{above}{wave}{Acoustic pressure}
\vspace{3em}

\begin{equation}
    \eqnmarkbox[Green]{ic1}{p(x,t=0) = p_0(x)}, \quad
     \eqnmarkbox[Green]{ic2}{\partial_t p(x,t=0) = 0},\\
\end{equation}
\annotate[yshift=1em]{above}{ic1}{Acoustic pressure at time 0}
\annotate[yshift=-0.5em]{below}{ic2}{At time 0, the pressure waves are not moving}
\vspace{3em}

\begin{equation}
    \eqnmarkbox[RedOrange]{meas}{g_n(x_n,t)} = 
    \eqnmarkbox[Plum]{mop}{\mathcal{M}_n p(x,t)} + 
    \eqnmarkbox[WildStrawberry]{noise}{\varepsilon_n(t)}, \quad 
    \eqnmarkbox[Cerulean]{ndet}{n = 1,\ldots,N}
\end{equation}
\annotate[yshift=1em]{above}{meas}{Boundary measurements}
\annotate[yshift=-0.5em]{below}{mop}{Sampling by the detection elements}
\annotate[yshift=1em]{above}{noise}{Noise}
\annotate[yshift=1em]{above}{ndet}{Number of detector elements}
\vspace{2em}

\noindent where $p(x,t)$ is the acoustic pressure in tissue with sound speed $v$ and mass density $\rho_0$. A variety of forms have been proposed for the absorption loss operator $\mathcal{L}$. The $N$ measured time series $g_n(x_n,t)$ are recorded at the $N$ detector points $x_n$ via the measurement operators $\mathcal{M}_n$. $\varepsilon_n$ represents noise. This mathematical model assumes the target is stationary on an acoustic time scale. However, if the measurements are not made simultaneously but sequentially with multiple excitation pulses, as is seen in some experimental systems, it might be that each measurement $g_n$ is generated by a slightly different initial pressure distribution $p_{0,n}(x)$, e.g.\ if the tissue moves between laser excitation pulses.
With the notation introduced here, many of the various assumptions typically made during image reconstruction can be written succinctly: spatial invariance of the tissue properties, e.g.\ $\nabla v = 0$, $\nabla \rho_0 = 0$, $\mathcal{L} = 0$, $\nabla \Gamma = 0$, or fluence approximations, e.g.\ $\nabla \Phi = 0$, or $\Phi(x) = \exp(-ax)$, or no tissue motion
$p_{0,n} = p_0$, or assumptions about the detection process, such as that $\mathcal{M}$ is omnidirectional or has a flat frequency response.

\section{Artifacts in PAI}
The following sections will discuss artifacts encountered in PAI, grouped by source (Fig.~\ref{fig:overview}). Each artifact will be explained and illustrated by a figure that demonstrates its origin through simulations, accompanied by experimental data whenever possible. The simulations are performed using SIMPA~\cite{grohl2022simpa}, see \url{https://gitlab.com/MTRietberg/artifact-paper} for our code.

\begin{figure}
    \centering
    \includegraphics[scale=0.37]{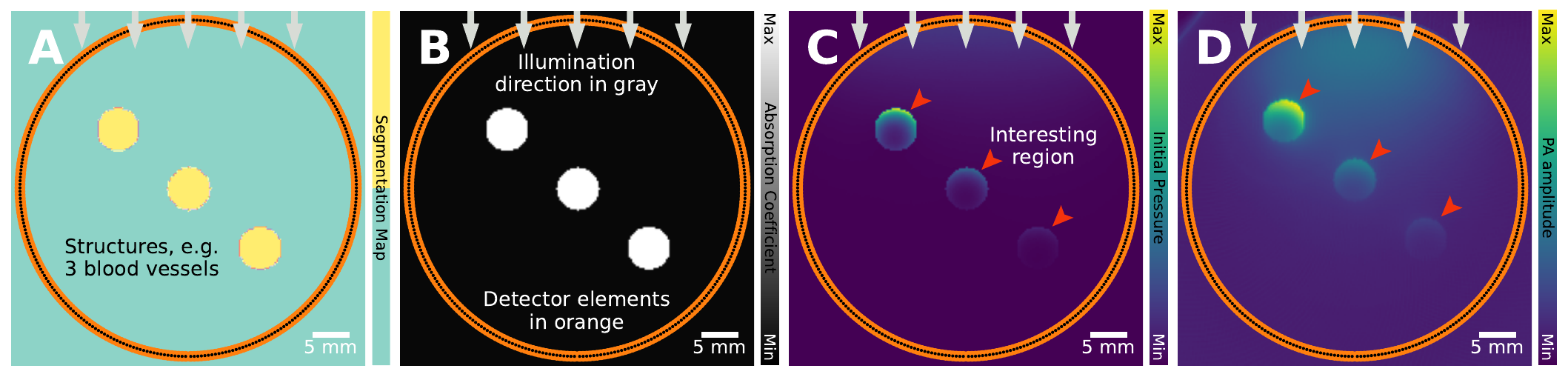}
    \caption{\textbf{A visual guide of how we will showcase visual explanations of the artifacts.} In this example, three vessels are used and we typically showcase different types of views. \textbf{A}: the segmentation map (a segmented color bar shows the considered tissue types). \textbf{B}: the absorption coefficient in a gray-scale color map. \textbf{C}: the initial pressure and in \textbf{D}: the reconstructed image, both with a pseudo-colored visualization using the \textit{Viridis} color mapping. In other figures, only the relevant components will be shown, e.g.\ only the absorption coefficient and the reconstructed image. In all figures, the illumination direction is denoted with gray arrows, the detector elements in orange (which appear in this figure as a continuous line due to the high detector density), and the scale with a scalebar at the bottom right corner. Noteworthy regions are indicated with red arrowheads. In addition to these plots, we also show reprinted examples from the literature, where possible.}
    \label{fig:example}
\end{figure}

In order to isolate particular artifacts, the simulations are (unless otherwise noted) performed in ideal settings, but for the one difference giving rise to the artifact. The ideal settings are: no patient movement, short enough laser pulse (satisfying the thermal and stress confinements for the imaged structures), no pulse-to-pulse laser power variation (50 mJ), all medium properties homogeneous (constant, the same everywhere), no acoustic attenuation, complete data collected by enclosing the region-of-interest with a ring-shaped transducer array with a high detector element density, detector elements with omnidirectional and broadband sensitivity, no added noise in both initial pressure and time series, and sufficient temporal sampling (40 MHz). The only artifact cause that is consistently present throughout all simulations is fluence decay, as the resulting artifact is unavoidably present in all PA images. The optical part of the simulation is performed in 3D (three spatial dimensions) and the acoustic part in 2D, with the exception of the out-of-plane absorption artifact (where both the optical and the acoustic parts are performed in 3D). The volume is 50x50x50 mm, with a resolution of 0.25 mm/pixel. The default delay and sum beamforming algorithm from SIMPA is used~\cite{jimaging4100121}, as delay and sum is widely used and relatively simple. Reconstruction techniques can incur artifacts themselves, e.g.\ time reversal can trap artifacts that would otherwise lie outside the image region~\cite{cox2009artifact}, but discussion of these artifacts is outside the scope of this paper. An example figure with labels is shown in Fig.~\ref{fig:example}.

\section{Artifact Source: Light-Tissue Interactions}
The interactions of light with tissue form the first part of the multi-physics process of photoacoustic signal generation, where several artifacts have their origin. Light scatters strongly in biological tissues and photons stochastically get absorbed along their travel paths, leading to signal changes. Propagation of photons in three spatial dimensions also leads to pressure buildup outside of the imaging plane. Furthermore, changes in the laser intensity will change the intensity of the PA image, and finally, the PA effect, which is governed by thermal and stress confinement conditions, and the duration of the light pulse can drastically change the appearance of the generated PA signal.

\subsection{Fluence Decay}
\label{sec:light:fluence-decay}
The pressure amplitude of the PA waves generated by a chromophore is directly proportional to both the absorption coefficient and the light fluence (Eq.~\ref{eq:absorbed_energy}). 
As the illumination beam propagates in tissue, the scattering and absorption of photons cause a subsequent decrease in the light fluence and thus PA signal intensity with depth (Eq.~\ref{eq:RTE_time}). Additionally, strong absorption by discrete structures can form so-called shadows behind them, resulting in a lower PA signal intensity. In the diffuse regime of photon travel, there are typically no hard shadows observed, but an exponential decrease in the light fluence with imaging depth (i.e. fluence decay) is visible. We can distinguish three qualitatively different effects of fluence decay: 
\begin{enumerate}
    \item The gradual decay of fluence across the photon travel distance (typically imaging depth).
    \item The rapid decay within a large absorbing object such that the center is less visible.
    \item Optical shadowing, where multiple absorbers lie close to each other and mutually influence the fluence.
\end{enumerate}

\begin{figure}[!htb]
    \centering
    \includegraphics[scale=0.37]{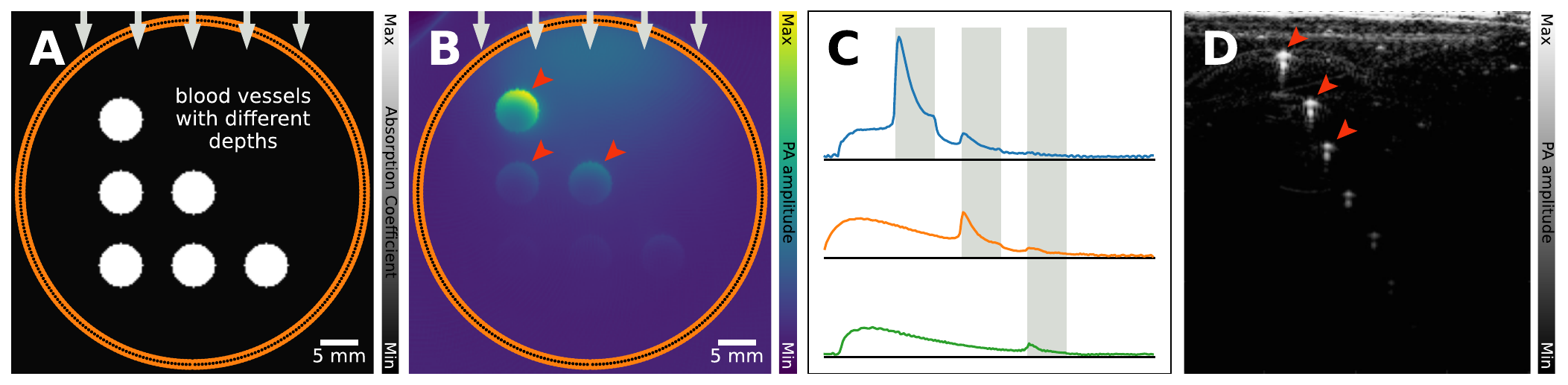}
    \caption{\textbf{Fluence decay artifact: leads signal loss with depth .} Several blood vessels are embedded in soft tissue, with increasing distance from the illumination source (increasing depth). \textbf{A}: their absorption coefficients. \textbf{B}: The fluence decays due to absorption from both the background and the blood vessels themselves, resulting in differing PA signal levels in the reconstruction. \textbf{C}: vertical intensity profiles taken from the reconstruction from \textbf{B}, showing fluence decay. This can also be seen in experimental data. \textbf{D}: bovine blood-filled tubes, embedded in polyvinyl chloride plastisol, measured with 800 nm. Just like our simulated example, tubes closer to the measurement surface result in a higher intensity than those deeper. The experimental image in \textbf{D} has been reprinted from Vogt et al., 2019 with permission from the publisher~\cite{Vogt:19}.}
    \label{fig:fluence_decay}
\end{figure}

We designed the example shown in Fig.~\ref{fig:fluence_decay} in a way that all three effects of this artifact can be seen in the reconstruction (Fig.~\ref{fig:fluence_decay}\textbf{B}). The vessel in the first row clearly shows the rapid decay of signal from the boundary to the core. In the second row, we see two vessels that have different signal amplitudes, caused by the asymmetrical absorption of photons of the vessel in the first row. The vessels in the third row are barely visible, highlighting the gradual decay of fluence with depth. 

Fluence decay is present in nearly all PA images, and if not corrected for (i.e., by assuming spatially invariant fluence), can lead to over- or under-estimation of PA signals. However, the correction of fluence decay is complicated due to the nonlinear behavior and ill-posed nature of the inverse problem~\cite{tarvainen2024quantitative}. A simple mitigation that is commonly applied in practice is to use an exponential gain to the image, either radially or with depth. The exponential gain can be based on estimated optical tissue properties, e.g., the effective attenuation coefficient $\mu_{\text{eff}}$, but will also amplify noise and thus not significantly increase the signal-to-noise ratio.

\subsection{Spectral Coloring}
\label{sec:light:spectral-coloring}
One can form PA images from a single excitation wavelength, but the true diagnostic power of PA lies within the possibility of multi-wavelength imaging. Performing multi-wavelength imaging provides a \textit{PA amplitude} spectrum for each spatial point in the image, which can (in theory) be used to extract molecular and functional information of the imaged tissue. During linear spectral unmixing to match the PA amplitude spectrum with the absorption coefficient spectrum of target molecules, it is typically assumed that a PA spectrum is proportional to the absorption coefficient spectrum at a point~\cite{grohl2024distribution}. Eq.~\ref{eq:initial_pressure}, however, shows that the PA amplitude depends on the local fluence, which is not constant inside tissue. Instead, the fluence suffers a decay in tissue (Sec.~\ref{sec:light:fluence-decay}). Furthermore, the scattering coefficient and especially the absorption coefficient are wavelength dependent, so that the fluence could experience marked reductions at selected spectral bands. The light reaching the structure is thus "colored" or filtered by the intervening tissue. Hence the name: spectral coloring. The PA spectrum from a certain location is then a distorted version of the correct spectrum expected from that location, i.e.: spectral coloring describes the preferential absorption of light at shorter wavelengths as a spectrum of light travels through the tissue

\begin{figure}[!htb]
    \centering
    \includegraphics[scale=0.37]{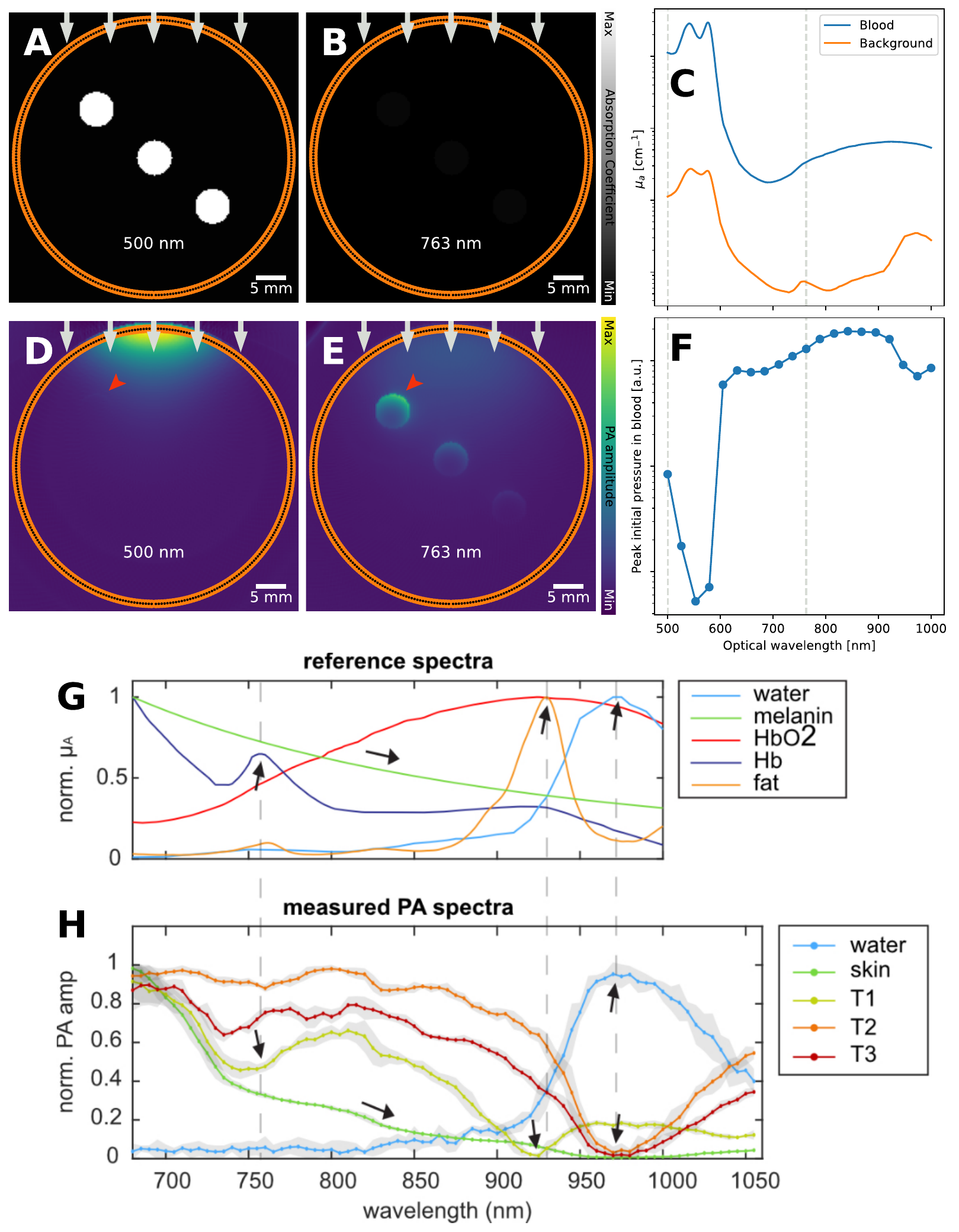}
    \caption{\textbf{Spectral coloring artifact: leads to unexpected signal difference between wavelengths.} 3 blood vessels are embedded in soft tissue, under identical settings, except the optical wavelength changes (\textbf{A}\&\textbf{D}: 500 nm, \textbf{B}\&\textbf{E}, 763 nm). \textbf{A}\&\textbf{B}: absorption coefficients. 
    \textbf{D}\&\textbf{E}: reconstruction, as an extension of the fluence decay artifact, the PA signal of the object under study is dependent on both the absorption spectrum of the object itself but also the background. In \textbf{D}, the blood vessels are not visible, as the light is being absorbed at the surface, but in \textbf{E}, the background absorption is weaker, meaning that the vessels are visible (even though the absorption coefficient of blood is lower than in \textbf{D}). \textbf{C}: the absorption coefficient against the optical wavelength (with the wavelengths from \textbf{A}\&\textbf{B} highlighted), and \textbf{F}: peak initial pressure against optical wavelength. This can also be seen in experimental data, such as the one shown in \textbf{G} and \textbf{H}. \textbf{G}: the reference absorption spectra, and \textbf{H}: the measured PA spectra of several locations around or in a healthy breast (water in the imaging bowl, T1, T2, T3 are points lying deeper in the breast). In \textbf{H} one can see the effect of the spectral coloring, e.g., T1 has a local minimum at 755 nm which corresponds to a peak in the reference spectrum of Hb. The \textit{in vivo} graphs in \textbf{G} and \textbf{H} have been reprinted from Dantuma et al., 2023~\cite{dantuma2023fullythreedimensionalsoundspeedcorrected}, which is available open-access under a CC BY 4.0 license.}
    \label{fig:spectral_coloring}
\end{figure}

Spectral coloring can be seen in Fig.~\ref{fig:spectral_coloring}. While at 500 nm the vessels have a high absorption coefficient (Fig.~\ref{fig:spectral_coloring}\textbf{A}\&\textbf{D}), they can only be resolved with 763 nm where their absorption coefficient is lower (Fig.~\ref{fig:spectral_coloring}\textbf{B}\&\textbf{E}). This difference can be attributed to the changing absorption coefficient of the background. This can also be seen in experimental results. The reference absorption spectra of water, melanin, HbO$_2$, Hb and fat differ a lot (Fig.~\ref{fig:spectral_coloring}\textbf{G}), which are chromophores expected to be present in the breast. We also see the spectral evolution of PA intensity in the RF signal from a single detector in the PAM3 breast imaging system (Fig.~\ref{fig:spectral_coloring}\textbf{H}). The PA intensities labeled `water', `skin' and `T1' correspond to parts of the RF signal which are dominantly coming from water, skin and inside the breast respectively (see inset for a schematic). It can be seen that while the PA spectrum for water matches the reference spectrum, the PA spectrum measured from skin shows the melanin reference spectrum modulated by absorptions coming from other chromophores. The effect is more significant in the PA spectrum `T1' where local dips in the spectrum match the absorption peaks of presumably the blood and fat contents of intervening tissues.  

Correction for spectral coloring can, in theory, be done by correcting for fluence decay at each optical wavelength, but it is very difficult to achieve in practice since accurate modeling of the fluence requires accurate knowledge of the optical properties in the first place. A more practical mitigation approach is solely looking at relative changes in a derived functional biomarker~\cite{tomaszewski2018oxygen}, on the assumption that the fluence will not significantly affect the relative change. Another approach is to restrict the analysis close to the tissue surface or to the pixels with maximum intensity projection (e.g.\ in Kirchner et al.~\cite{kirchner2018context}), under the assumption that uncolored light dominates these regions.

\subsection{Out-of-plane Absorption}
\label{sec:light:out-of-plane}
\begin{figure}[!htb]
    \centering
    \includegraphics[scale=0.37]{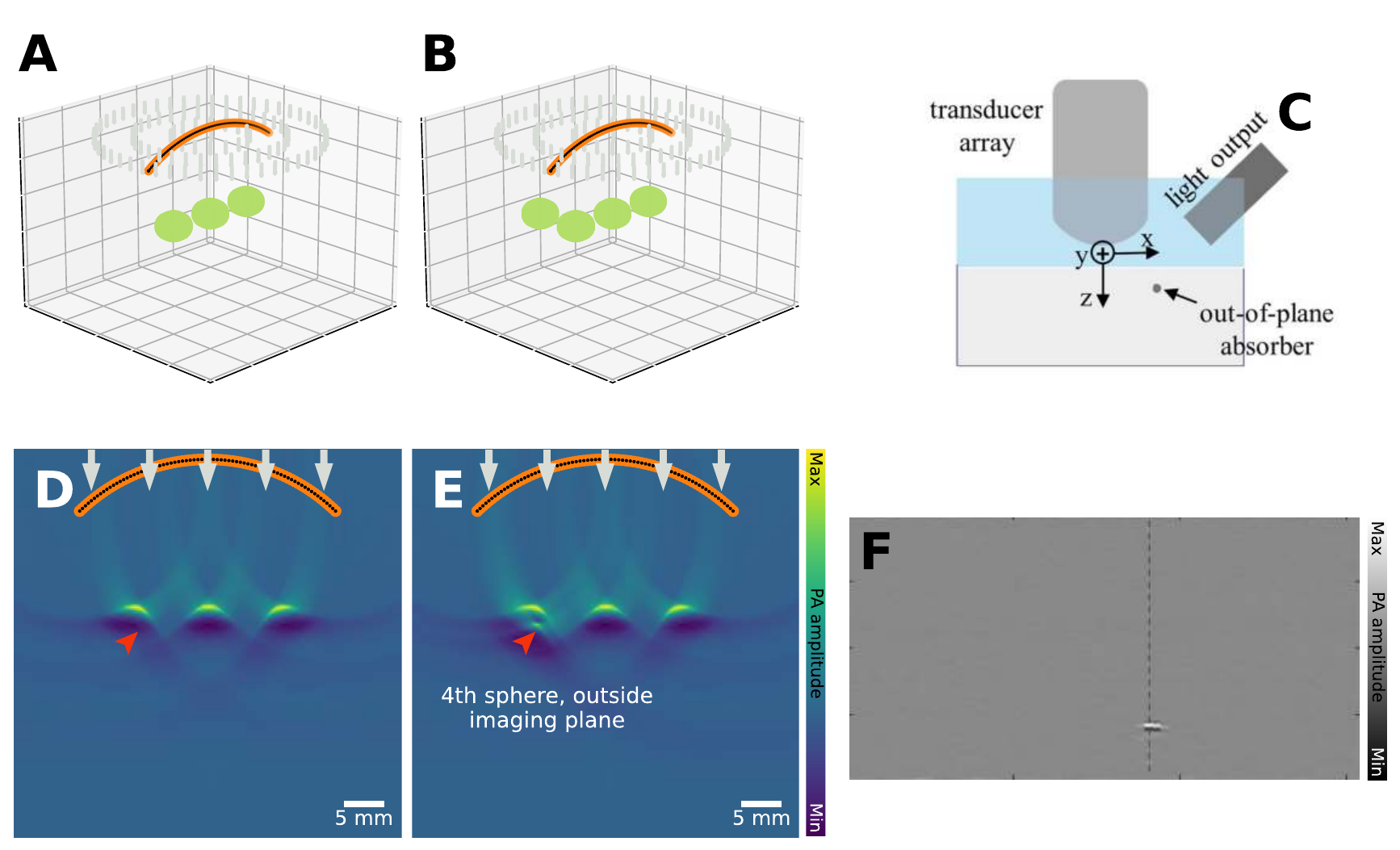}
    \caption{\textbf{Out-of-plane absorption artifact: leads to clutter.} The assumption is that only tissue in the imaging plane or volume is illuminated, as in the case in \textbf{A}\&\textbf{D}, where 3 spheres are imaged. The second column (\textbf{B}\&\textbf{E}) shows the same case, but now with a 4th object outside the imaging plane that is illuminated, which is reconstructed inside the imaging plane. \textbf{A}\&\textbf{B}: segmentation map, and \textbf{D}\&\textbf{E}: reconstruction. This can also be seen in experimental data, such as the one shown in \textbf{C}\&\textbf{F}. \textbf{C}: measurement configuration, where an absorber is placed out-of-plane (imaging plane is in Y-Z). \textbf{F}: reconstructed image, where the absorber is erroneously shown inside of the imaging plane. The experimental images have been reprinted from Nguyen et al., 2020~\cite{NGUYEN2020100176}, which is available open-access under a CC BY-NC-ND 4.0 license.}
    \label{fig:out_of_plane_illumination}
\end{figure}

In 2D PAI using e.g.\ a linear array or a ring array, it is often assumed that all detected PA signals originate from this plane, i.e., only absorbers in this slice of the tissue are excited. In practice, the highly optically scattering nature of tissue means that PA signals can be generated outside of the region of interest and be detected within the same time series. Detected signals from outside the imaging region are then projected inside the region. Fig.~\ref{fig:out_of_plane_illumination} shows where the addition of a sphere outside the imaging plane appears in the reconstructed image. Such artifacts can partly be corrected by changing from a 2D transducer array and reconstruction technique to 3D variants or increasing the directionality of the transducer elements. Furthermore, in real-time handheld PA devices, the situation could be ameliorated by the operator moving the image plane slightly to see that there is an absorber out-of-plane.

\subsection{Laser Power Variation}
\label{sec:light:laser-intensity}
\begin{figure}[!htb]
    \centering
    \includegraphics[scale=0.37]{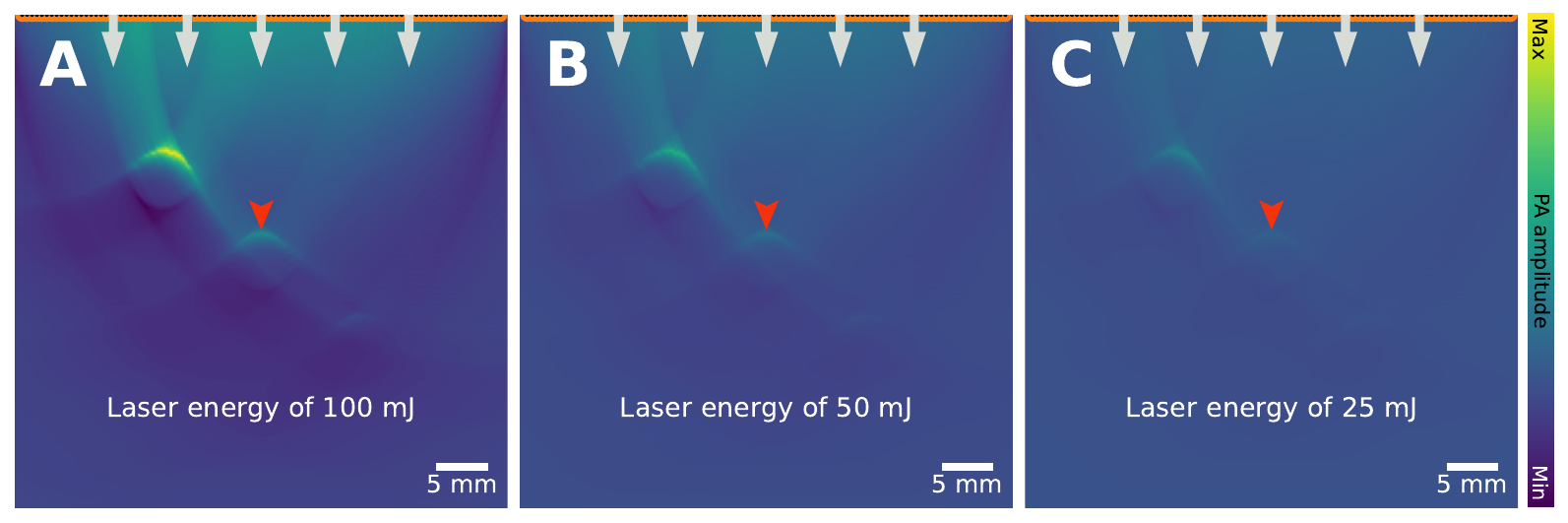}
    \caption{\textbf{Laser power variation artifact: leads to unexpected signal difference.} It is often assumed that the laser power is constant between measurements, which might not always be the case. Three blood vessels are measured in identical settings, except the laser power varies, resulting in change in PA signal. \textbf{A}: laser energy of 100 mJ, \textbf{B}: 50 mJ, \textbf{C}: 25 mJ, all at 5 ns pulse duration.}
    \label{fig:laser_power}
\end{figure}

In certain cases, a single measurement is not sufficient to create a PA image. For example, in raster-scanning systems (e.g.\ linear or rotating) and when making multi-wavelength measurements. When imaging a subject multiple times, the assumption is often implicitly made that all changes between frames originate from within the subject (e.g.\ heartbeat). However, if the laser pulse-to-pulse power is not stable over time, the resultant signal will change, even if everything else remains the same. If not corrected, laser power variations lead to global changes in the measured PA signal across illumination wavelength and time, hampering image quantification. Fortunately, correction of fluctuating laser power is routinely done via hardware solutions, and static wavelength-dependent changes can be calibrated for before the measurements. For live measurements of the laser power during acquisition, a beamsplitter is typically used to redirect a small percentage (e.g.\ <1\%) of the total light to another sensor. The artifact is visualized in Fig.~\ref{fig:laser_power}, where three blood vessels are imaged with varying laser power.

\subsection{Long Pulse Duration}
\label{sec:light:confinement}

\begin{figure}[!htb]
    \centering
    \includegraphics[scale=0.37]{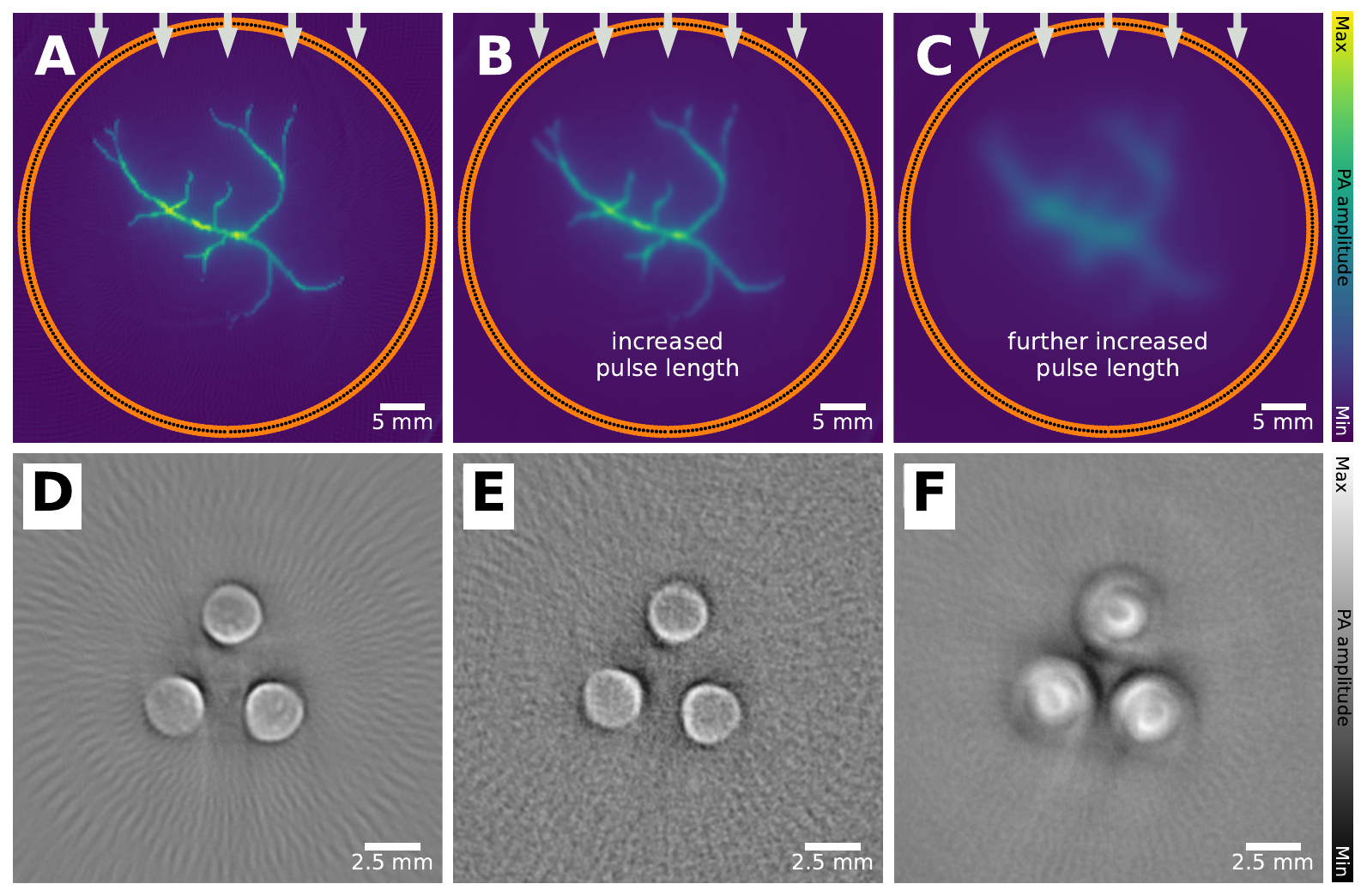}
    \caption{\textbf{Long pulse duration artifact: leads to blurring.} If the laser pulse is not short enough, confinement conditions are violated and PA signal is distorted. \textbf{A}: when imaging with a short enough pulse, the reconstruction is not distorted, \textbf{B}: increasing the pulse length will distort the reconstruction, \textbf{C}: further increases the pulse length will distort the image even more. This can also be seen in experimental data. \textbf{D--F} show measurements on three cylindrical absorbing elements inside a scattering gel, measured with \textbf{D}: 7, \textbf{E}: 65 and \textbf{F}: 500 ns pulse duration. The experimental images have been reprinted from Allen et al., 2006 with permission from the publisher~\cite{Allen:06}.}
    \label{fig:pulse_length}
\end{figure}

Illumination in PAI is most commonly achieved via a pulsed laser. The length of the laser pulses can be as low as femtoseconds~\cite{akhmanov1992optics}, but in the context of PAI these typically have a duration of 5 to 10 ns. These pulses should be short, such that the thermal and stress confinement conditions are fulfilled~\cite{wang2007biomedical}, and have sufficient pulse energy, while avoiding potential damage to tissue. Violating the confinement conditions with a longer pulse duration acts as a low-pass filter, meaning that no high acoustic frequencies are generated, resulting in blurry PA images, where features can be obscured. The effect of an extended duration pulse can be seen in Fig.~\ref{fig:pulse_length}, where an increasing pulse duration creates artifacts. Since the pulse length is a relatively stable property of the imaging system, pulse length artifacts are uncommon when imaging with commercially available, clinically approved systems. However, this artifact may become noticeable with the increasing uptake of light-emitting diode and laser diode illumination systems for clinical devices, which require longer pulses to compensate for their lower output power~\cite{Ozsoy2021,Liu2023}. Mitigation approaches used are often deconvolution-based deblurring~\cite{Rejesh:13}, which uses the fact that the degraded, detected signal is the result of a time convolution with the ideal PA signal and the laser pulse.

\clearpage % force Latex to drop any held images here, so we can keep the sections separate while editing.

\section{Artifact Source: The Photoacoustic Effect}

% \subsection{PA Efficiency}
The PA efficiency $\Gamma$ (Eq.~\ref{eq:initial_pressure}) is a material property that quantifies the pressure increase, $p_0$, resulting from the deposited thermal energy, originating from light absorption and non-radiative de-excitation~\cite{sigrist1986laser, wang2007biomedical}. The PA efficiency is defined for the medium where light absorption and pressure wave generation occur. In most biological tissues, blood dominates PA signal generation due to the strong optical absorption of hemoglobin in red blood cells. In such cases, under the confinement regimes the PA efficiency is primarily a property of blood rather than the surrounding tissue.

\begin{figure}[!htb]
    \centering
    \includegraphics[scale=0.37]{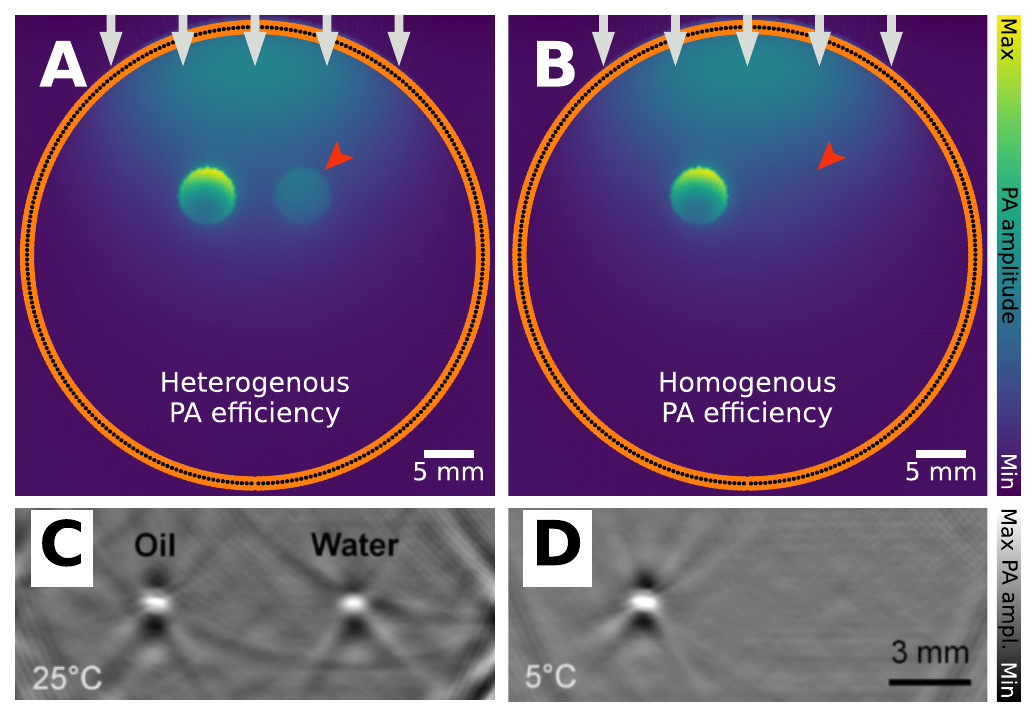}
    \caption{\textbf{PA efficiency artifact: leads to unexpected signal difference.} Often, PA efficiency is assumed to be homogenous, but this may not always be the case. \textbf{A}: a PA image of a fat blob (left) and blood vessel (right) with heterogeneous PA efficiency coefficient (0.8 for fat, 0.2004 for blood and background), \textbf{B}: a PA image with homogenous, 0.2004, PA efficiency coefficient. There is a clear difference in intensity, which is not the result of a changing absorption coefficient, but of a changing PA efficiency. This can also be seen in experimental data, where two tubes filled with mineral oil and pure water are imaged with PAI, imaged at \textbf{C}: 25\degree{}C and \textbf{D}: 5 \degree{}C. The change in temperature results in a difference in PA efficiency. The experimental images have been reprinted from Petrova et al., 2017~\cite{PETROVA201736}, which is available open-access under a CC BY-NC-ND 4.0 license.}
    \label{fig:PA_eff}
\end{figure}

PA signals can also originate from tissue types other than blood, if they have sufficiently high optical absorption at the illumination wavelength~\cite{diot2017multispectral}. When non-blood tissue components directly absorb light, their own PA efficiency governs the PA signal. Organs are inherently heterogeneous, consisting of different tissues such as muscle, adipose tissue, connective tissue, tumors, and fluid-filled lesions. Even though these different tissues clearly have varied molecular composition, it is often implicitly assumed that the PA efficiency is spatially uniform within an organ or tissue region. This assumption can lead to inaccuracies and artifacts in quantitative PAI, as the PA signal strength depends not only on the absorbed energy density but also on spatial variations in the PA efficiency, which can be seen in Fig.~\ref{fig:PA_eff}.

An often overlooked factor is that the PA efficiency is temperature-dependent~\cite{wang2007biomedical, daoudi2013two}. The PA efficiency may thus not be constant under different physiological or experimental conditions, which can introduce additional uncertainty where temperature effects are significant and has in fact shown promise in applications for thermal therapy monitoring with PAI~\cite{assi2020real, kim2019real, petrova2018vivo, zhou2019thermal}. The PA efficiency can also be concentration-dependent, although this typically only manifests in a problematic sense in copper and nickel sulfate solutions used in phantoms~\cite{fonseca2017sulfates}.

Assumptions regarding spatially uniform PA efficiency can be further confounded by the use of nanoparticle-based PA contrast agents, whereby the acoustic wave generation arises due to the rapid transfer of heat from the nanoparticles to the surrounding fluid: an \textit{indirect photoacoustic effect}.
In such a setting, the absorption coefficient, mass density and specific heat capacity would be those of the particles, but the thermal expansion coefficient and the isothermal compressibility would be those of the surrounding medium. In other words, the nanoparticle serves as a local heater, but the thermoelastic conversion into a pressure wave occurs in the surrounding medium~\cite{calasso2001photoacoustic, chen2011environment,pang2016photoacoustic,triki2018mathematical}. This phenomenon could introduce complexities in quantitative PAI, as the measured signal depends not solely on the imaging target (the nanoparticle).

\clearpage % force Latex to drop any held images here, so we can keep the sections separate while editing.

\section{Artifact Source: Sound-Tissue Interactions}
When reconstructing PA images from measured data, it is necessary to know the acoustic properties of the medium, in particular the sound speed. Although the acoustic properties of biological tissue usually vary across the field of view, but in general, a map of the acoustic properties is not available. For this reason, and because for soft tissue it is a reasonable starting assumption, used widely in ultrasound, the acoustic properties are commonly treated as though they are spatially homogeneous, and each is defined by a single value. Although this can be a good approximation, especially for small regions of soft tissue, when it fails, artifacts appear in the PA images as a consequence.

\subsection{Sound Speed}
\begin{figure}[!htb]
    \centering
    \includegraphics[scale=0.37]{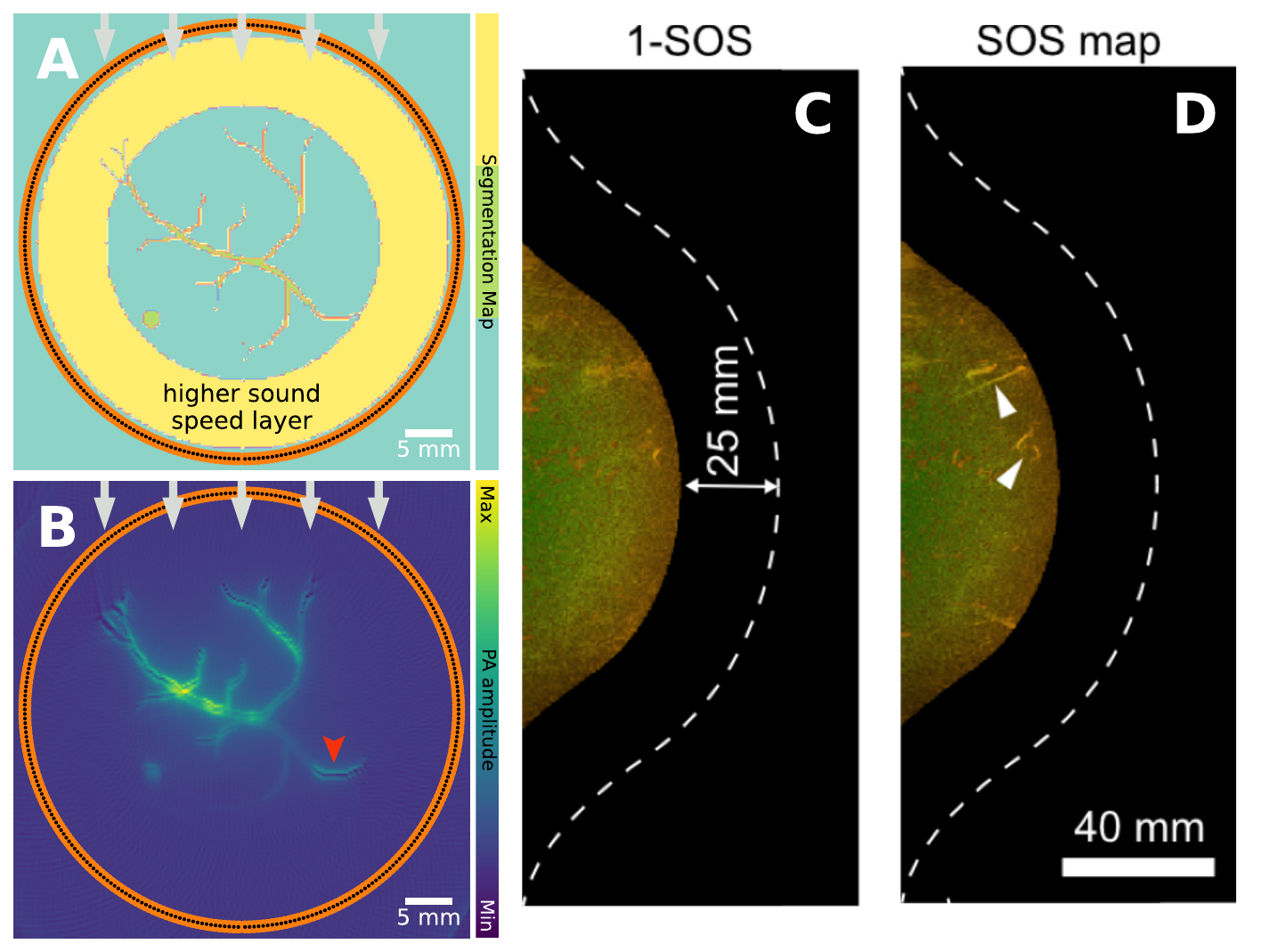}
    \caption{\textbf{Sound speed artifact: leads to dislocation, splitting, and blurring.} \textbf{A}\&\textbf{B}: Simulation showing the effect of a sound speed heterogeneity when imaging a vascular network assuming constant background sound speed.
    \textbf{A}: Schematic of the vascular target with background sound speed 1540 m/s, yellow annulus 1580 m/s.
    \textbf{B}: Image corresponding to \textbf{A} showing splitting and blurring artifacts due to the time delays introduced by the region of higher sound speed.    
    \textbf{C}\&\textbf{D}: Maximum intensity projections of \textit{in vivo} PA breast images, \textbf{C}: reconstructed assuming constant sound speed, and \textbf{D}: using a 3D measurement of the sound speed distribution obtained using ultrasound tomography. In both cases, the outermost 25 mm of tissue is not shown so that deeper vessels are visible and not obscured by more superficial vessels. The white dashed line marks the surface of the breast. The arrowheads indicate vascular structures that are are lost under the assumption that the sound speed is constant. The \textit{in vivo} images are reprinted from Dantuma et al., 2023~\cite{dantuma2023fullythreedimensionalsoundspeedcorrected}, under a CC BY 4.0 license).}
    \label{fig:sound_speed_heterogeneities}
\end{figure}

Accurate PA image reconstruction relies on being able to map measured time series back into the spatial domain through knowledge of the sound speed. To help visualize how inaccuracy in the sound speed can affect PA images, consider the simple case of a point-like PA source that emits a spherical pulse of sound in a medium with a constant sound speed $v$. For the simple case of a point-like PA source, variations in sound speed will impact the pulse of sound emitted isotropically, and due to interactions such as reflection, diffraction, and scattering, the pulse will also not necessarily travel along a straight path between the source and detector. In other words, sound speed heterogeneities can change the time it takes acoustic waves to reach the detectors as well as bend or scatter them. When the image reconstruction algorithm neglects these effects by assuming a homogeneous sound speed, artifacts will result, which can be seen in Fig.~\ref{fig:sound_speed_heterogeneities}. 

In the simulated examples (Fig.~\ref{fig:sound_speed_heterogeneities}\textbf{A}\&\textbf{B}), a vascular network is imaged under the assumption that the sound speed is constant when it contains a region with slightly different sound speed surrounding the target region, reflecting an imaging scenario in which the object (such as a breast or small animal) is surrounded by water as a coupling medium. Ignoring the difference in the sound speed can lead to characteristic splitting and blurring artifacts.
This is also visible in the experimental example (Fig.~\ref{fig:sound_speed_heterogeneities}\textbf{C}\&\textbf{D}), an \textit{in vivo} example of breast imaging~\cite{dantuma2023fullythreedimensionalsoundspeedcorrected}, in which two deep-lying vessels, which are visible in the image when the spatial variation in the sound speed distribution is taken into account, are not visible when the image is reconstructed assuming a constant sound speed.

When the detection array surrounds the object (Fig.~\ref{fig:sound_speed_heterogeneities}\textbf{A--D}), the effect on the image of the distortions that the PA waves accumulate as they travel across the whole imaged region can be mitigated by reconstructing the image using only the first half of the time series data, up to time $R/v$, where $R$ is the radius of the detection array~\cite{anastasio2005half}.
However, this is not possible for many PA scanner geometries. A useful but non-optimal approach to mitigating a non-uniform sound speed is to continue to assume a homogeneous sound speed but to optimize the value of the sound speed in some way~\cite{treeby2011automatic,yoon2012enhancement,cong2015photoacoustic,zhang2022video}.

\subsection{Acoustic Reflections}
Heterogeneous speed of sound or density across the field of view can affect image reconstruction in two ways: firstly, scattering can attenuate the PA signals, and secondly, waves scattered from the primary PA waves may themselves be of sufficient amplitude to be detected and thereby result in clutter artifacts in the images. Clutter is more likely when the scatterer is large compared with the PA pulse. One important case is when the primary PA waves are generated in the skin through absorption by melanin~\cite{else2024effects, Else2025.03.28.25324605}. Clutter has also been reported from PA waves generated at the face of the detector array itself~\cite{singh2015photoacoustic,Preisser2016}.

\begin{figure}[!htb]
    \centering
    \includegraphics[scale=0.37]{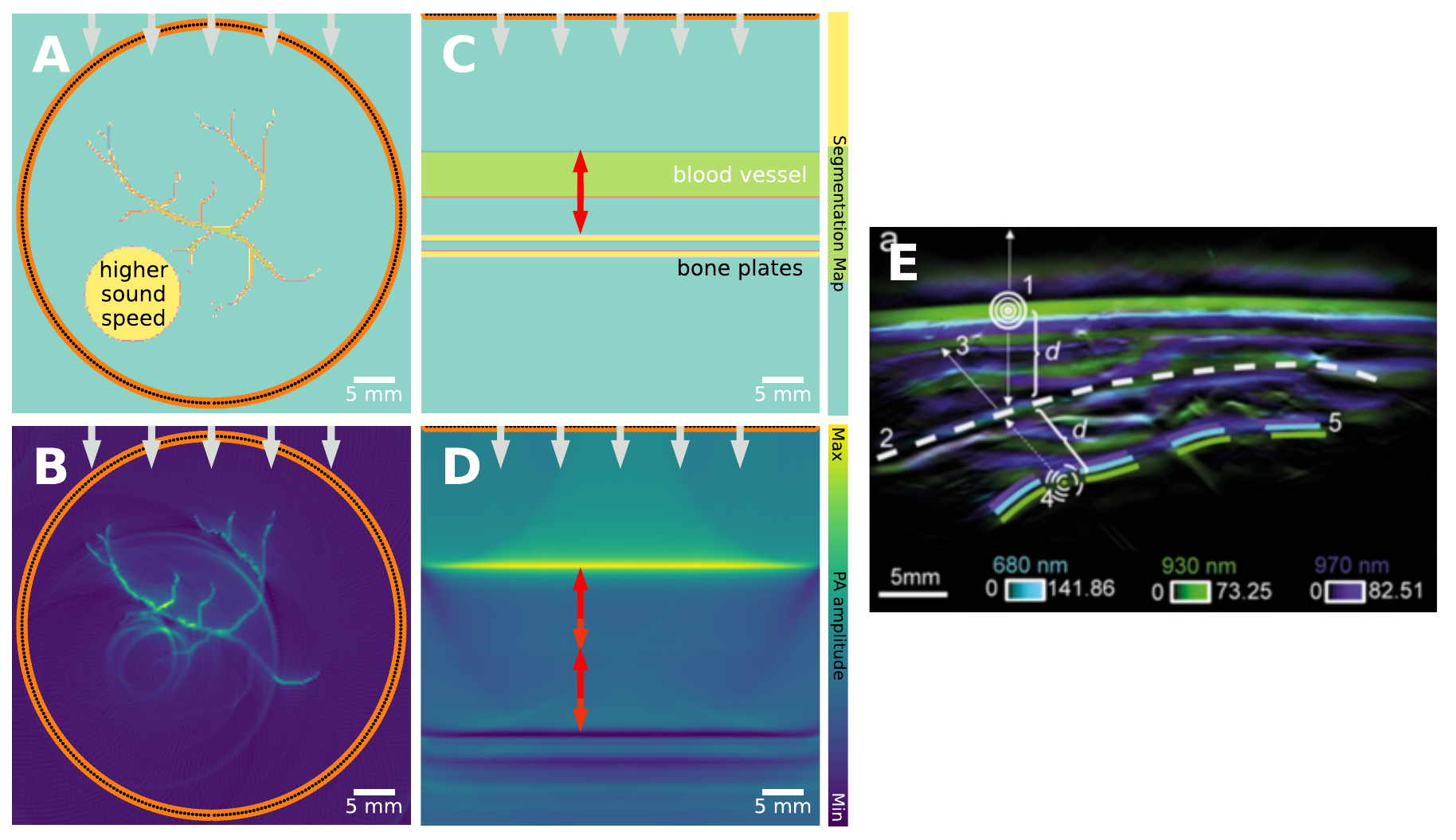}
    \caption{\textbf{Acoustic reflections artifact: leads to clutter.} 
    \textbf{A}: a schematic of a vascular target with a background sound speed of 1540 m/s and a heterogeneity (yellow disc, 1800 m/s).
    \textbf{B}: the image corresponding to \textbf{A}, reconstructed assuming a constant sound speed and therefore showing clutter (scattering) artifacts due to the reflections from the region of higher sound speed.
    \textbf{C}: a schematic of a single vessel imaged near a bone. \textbf{D}: the image corresponding to \textbf{C}, reconstructed assuming a constant sound speed, and the reflections from the bone are clearly visible at twice the blood-bone distance (red arrow).     
    \textbf{E}: \textit{in vivo} image~\cite{noltes2023towards} 
    of a papillary thyroid carcinoma, delineated in white. The PA image exhibits reflection artifacts (dashed line 5) of the skin signal caused by the capsule of the thyroid nodule (dashed line 2). PA signals originating in point 1 are reflected at the capsule (arrow 3), and an artificial source (point 4) is mirrored to the opposite side.
    The \textit{in vivo} image is reprinted from Noltes et al., 2023~\cite{noltes2023towards}, under a CC BY 4.0 license.}
    \label{fig:acoustic_reflections}
\end{figure}

This artifact is demonstrated in Fig.~\ref{fig:acoustic_reflections}. A vascular target is imaged through a medium with background sound speed 1540 m/s and a circular region with high sound speed of 1800 m/s (Fig.~\ref{fig:acoustic_reflections}\textbf{A}\&\textbf{B}). Here, the heterogeneous region acts as a strong scatterer, and the clutter artifacts that result from this scattered part of the signal not being mapped back into the correct region in the image are clearly visible. In a different case, the scatterer is a bone layer lying parallel to the vessel of interest (Fig.~\ref{fig:acoustic_reflections}\textbf{C}\&\textbf{D}). Here, the reflection from the bone appears in the image as another linear feature at twice the blood-bone distance. Reflection artifacts are also visible in the \textit{in vivo} example (Fig.~\ref{fig:acoustic_reflections}\textbf{E}), showing reflection artifacts in an image of a papillary thyroid carcinoma

In some simple cases, it may be possible to mitigate the presence of reflected signals by time-gating in the raw time series. For example, this would be possible in the linear example (Fig.~\ref{fig:acoustic_reflections}\textbf{C}\&\textbf{D}) as the reflections will arrive later than the primary PA signal.

\subsection{Acoustic Attenuation}
\label{subsec:attenuation}
\begin{figure}[!htb]
    \centering
    \includegraphics[scale=0.37]{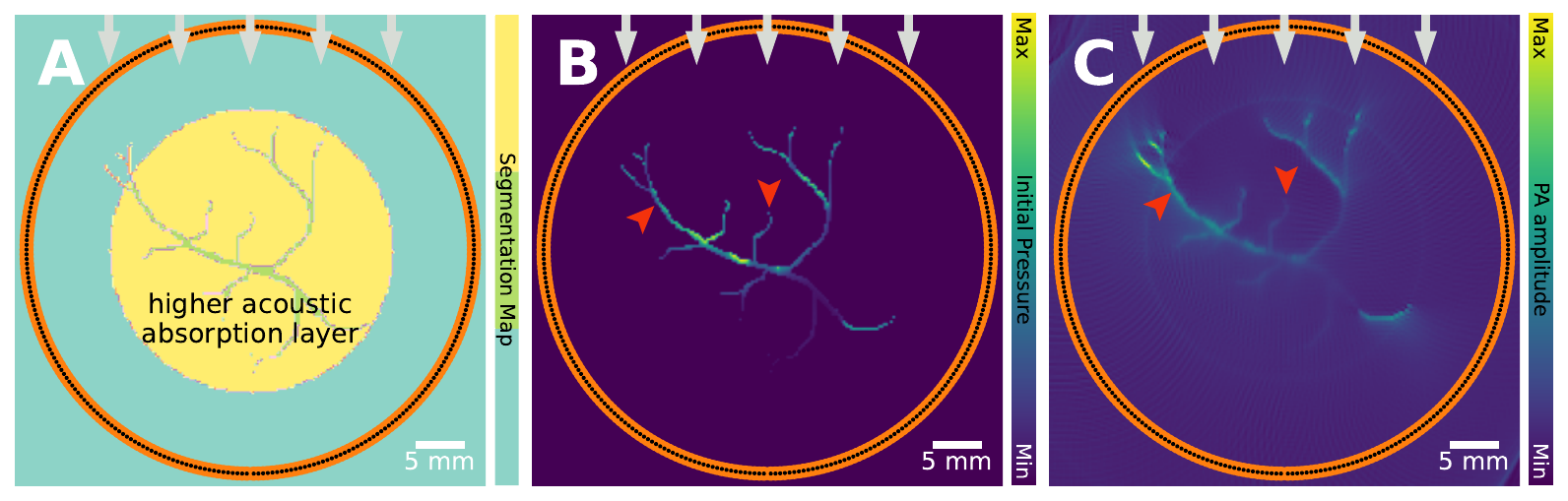}
    \caption{\textbf{Acoustic attenuation artifact: leads signal loss with depth.} 
    \textbf{A}: Schematic of a vascular network in tissue which is homogeneous except for a region with higher absorption (yellow).
    \textbf{B}: Initial pressure, i.e. before acoustic absorption,    \textbf{C}: Image reconstructed while there is absorption in the medium (yellow region), but the reconstruction assumes there is none. The waves that have traveled through the most tissue (those traveling from the image center) are most strongly absorbed. As absorption is stronger at higher frequencies, the images are most blurred near the center (compare the two vessels pointed to by the arrowheads, one near the center, one near the edge of the absorbing region).}
    \label{fig:acoustic_attenuation}
\end{figure}

As PA waves travel through tissue, they are scattered to some extent. While scattering from large scatterers can result in discrete measurable reflections, as discussed above, the main effect of scattering from small scatterers (diffuse or Rayleigh scattering) is to attenuate the primary PA waves, an effect that increases strongly as a function of frequency. In addition, acoustic absorption (conversion of acoustic energy to heat) will occur in biological tissues and also increases strongly with frequency. Attenuation (an umbrella term used to include the effects of both scattering and absorption) therefore decreases the PA signal amplitude, preferentially removing the higher frequencies. The result is a decrease in the image resolution since ultrasound frequency defines the axial and lateral resolution of the image.
This can be seen in Fig.~\ref{fig:acoustic_attenuation}, in which the presence of an absorbing region attenuates the signals traveling through it. As absorption in tissue more strongly attenuates higher frequencies, the waves from the center of the image, which have traveled furthest, have the highest frequency content removed, hence vessels near the center of the image are blurred.
The effect of acoustic attenuation can be ameliorated through deconvolution to some extent if its frequency-dependence is known~\cite{treeby2013acoustic}, although when the signal has fallen into the measurement noise, it cannot be recovered.

\clearpage % force Latex to drop any held images here, so we can keep the sections separate while editing.

\section{Artifact Source: Signal Detection}
The previous sections discussed sources of artifacts caused by the difference between the true physical interactions of light and sound with tissue and the assumptions made in the image reconstruction. This section will discuss limitations in the acoustic signal detection hardware that can lead to an insufficient capture of PA pressure waves, introducing artifacts into the reconstructed images because of a lack of data. These artifacts arise from: (1) sampling of the acoustic field, as we can only have discrete measurement points, which are subject to the arrangement and pose of the transducer elements in 3D space and temporal sampling of the acoustic pressure, (2) the response of the detector elements, as acoustic detectors respond differently to the incoming acoustic pressure depending on the frequency of the waves and the incident angle, and (3) noise, as measurements are subject to electrical and thermal effects.

\subsection{Sampling of the Acoustic Field}
\begin{figure}[!htb]
    \includegraphics[width=0.95\linewidth]{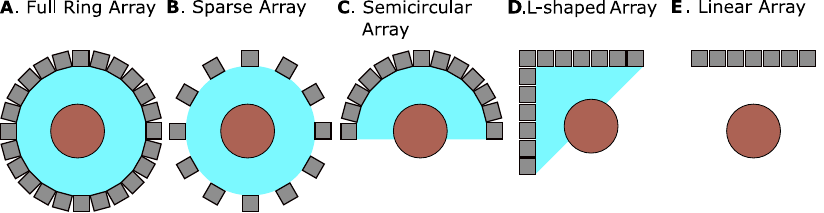}\\
    \vspace*{0.3em}
    \centering
    \includegraphics[scale=0.37]{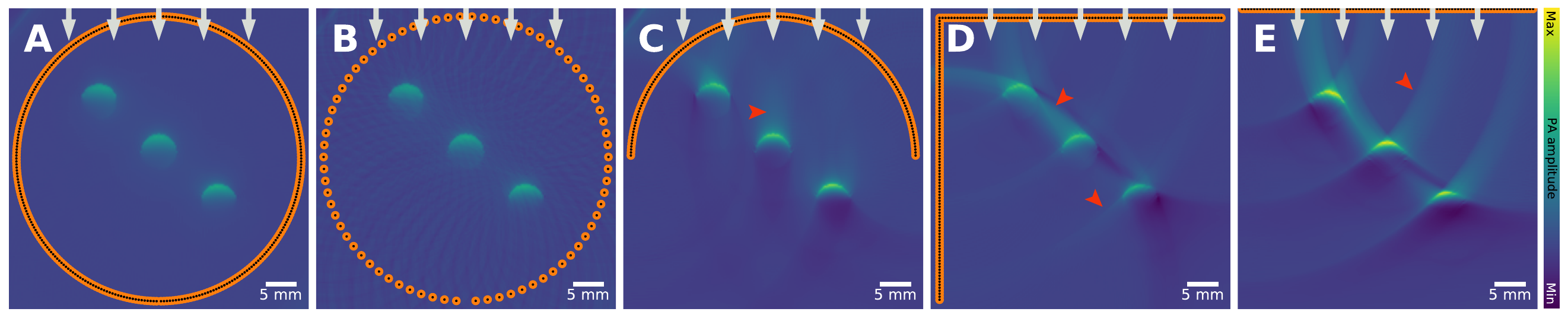}
    \caption{\textbf{Example visualisation of various typical detector configurations for PAI devices.} Three blood vessels are shown and images are reconstructed from simulations of the following detector configurations: \textbf{A}: a full-ring array, \textbf{B}: a sparse full-ring array, \textbf{C}: a semi-circular array, \textbf{D}: an L-shaped array, and \textbf{E}: a linear array. The upper row shows schematics, including the \textit{visible region} of the image highlighted in light blue. The lower row shows simulations of the same initial pressure distribution but measured and reconstructed with different detection geometries.}
    \label{fig:pa_device_configurations}
\end{figure}

PAI scanners have been implemented in various configurations, catering to the application's requirements~\cite{yang2022practical}. Pre-clinical PA scanners for small animal imaging, can typically be found in a tomographic imaging setting, optimized for slice-wise imaging through the specimen with full angular coverage (Fig.~\ref{fig:pa_device_configurations}\textbf{A}, e.g.~\cite{mervcep2019transmission}), sometimes with a sparsely populated arrays (Fig.~\ref{fig:pa_device_configurations}\textbf{B}, e.g.~\cite{guan2020limited}), or with reduced angular coverage (Fig.~\ref{fig:pa_device_configurations}\textbf{C}, e.g.~\cite{joseph2017evaluation}). Most clinical PAI scanners, on the other hand, are operated in a handheld mode, similar to conventional ultrasound scanners. These record 2D image slices and have either limited angular coverage (Fig.~\ref{fig:pa_device_configurations}\textbf{C}, e.g.~\cite{tan2024non}) or are even used with a linear detection array (Fig.~\ref{fig:pa_device_configurations}\textbf{E}, e.g.~\cite{menezes2018downgrading}). There exist scanners with other geometries as well. One example is an L-shaped detector (Fig.~\ref{fig:pa_device_configurations}\textbf{D}, e.g.~\cite{ellwood2017photoacoustic}) that can reconstruct a greater field of view by placing a second linear array orthogonal to the first. Another example is a three-dimensional hemispherical detector array that can be used for breast imaging~\cite{schoustra2023imaging}, or 3D tomographic preclinical imaging~\cite{kalva2023spiral}. For endoscopic~\cite{guo2020photoacoustic}, mesoscopic~\cite{hacker2023performance}, or microscopic~\cite{yao2013photoacoustic} PAI, single detectors are often used. When using a single detection element, images can be reconstructed by mechanical scanning (translating and rotating a singular detector in space) or by optical scanning (deflecting the excitation laser across the tissue).

\subsubsection{Limited View}

\begin{figure}[!htb]
    \centering
    \includegraphics[scale=0.37]{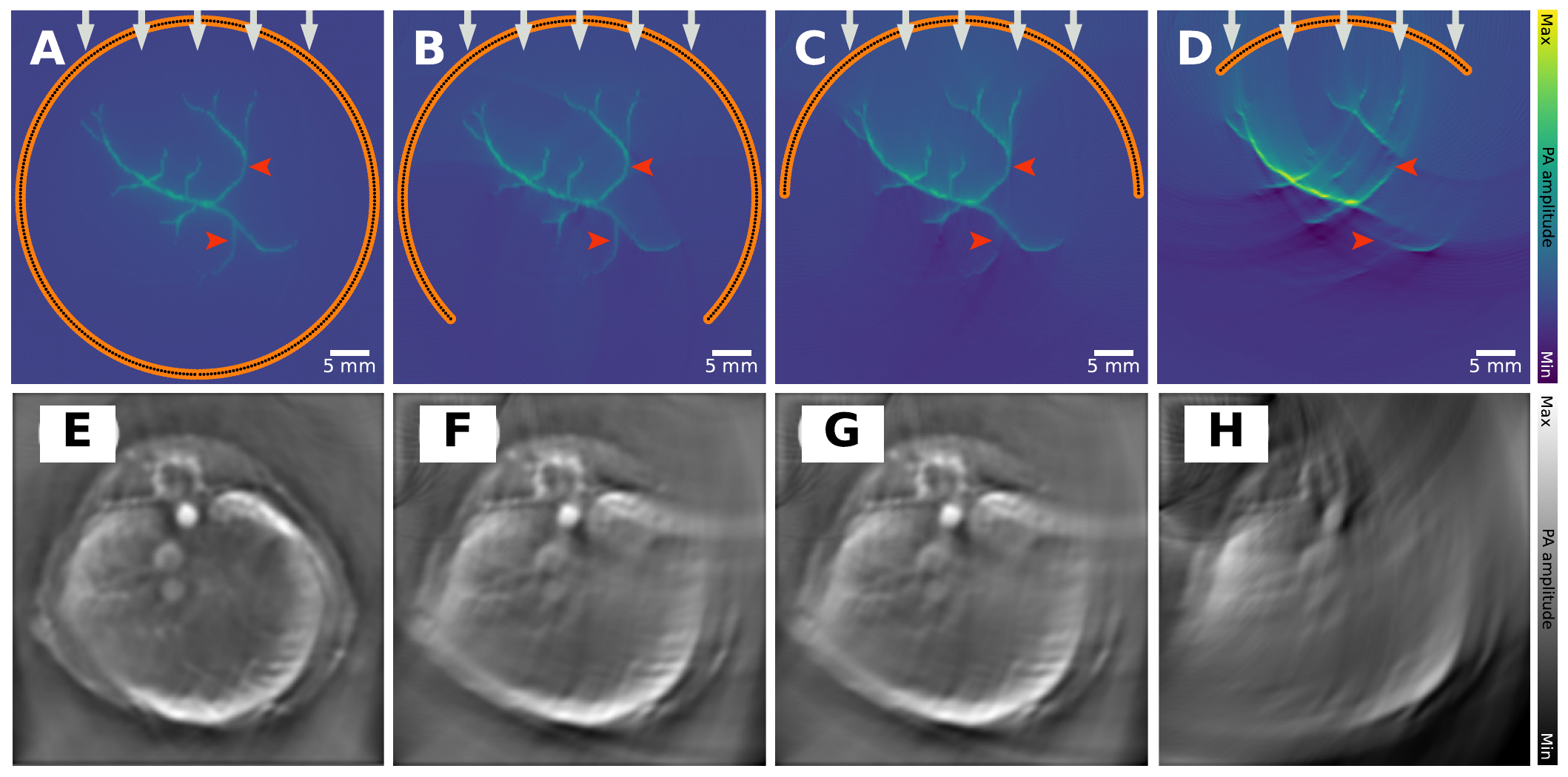}
    \caption{\textbf{Limited view artifact: leads to clutter and signal loss.} A vascular network is simulated under identical settings, except the detector opening angle of a spherical detector changes. As the detection region (the region that is enclosed by the detectors) becomes smaller (\textbf{A}: 360\degree, \textbf{B}: 270\degree, \textbf{C}: 180\degree, \textbf{D}: 90\degree) the image quality increasingly drops. We show the limited view artifacts for simulations (\textbf{A--D}) as well as preclinical \textit{in vivo} data, in this case of a mouse (\textbf{E--H}). \textbf{E}: 180\degree, \textbf{F}: 120\degree, \textbf{G}: 90\degree and \textbf{H}: 70\degree. The \textit{in vivo} images have been reprinted from Xiao et al., 2025~\cite{xiao2025limited}, which is available open-access under a CC BY-NC-ND 4.0 license.}
    \label{fig:hardware:limited_view}
\end{figure}

In the limited view problem, a \textit{visible region} can be defined, described by the envelope of the detection curve (2D) or surface (3D). This visible region is indicated in blue in Fig.~\ref{fig:pa_device_configurations}. In a 3D setting, any object in the visible region can (in theory) be reconstructed accurately, and in 2D, its edges can be reconstructed accurately~\cite{xu2004reconstructions,paltauf2007experimental}. On the other hand, if the imaging target is outside of the visible region, it leads to characteristic artifacts that depend on the detection geometry and the employed reconstruction algorithm in Fig.~\ref{fig:hardware:limited_view}.

For situations in which an increase in the angular coverage around the region of interest is not feasible, several hardware- and software-based mitigation strategies have been proposed to date. Hardware-based approaches include the addition of an acoustic reflector to form an additional virtual array to effectively double the detection view~\cite{huang2013improving}, or the use of curved linear array transducers~\cite{choi2018clinical}. Computational approaches to mitigate limited view artifacts include specialised modifications to the reconstruction algorithm, such as the inclusion of adaptive weighting for filtered backprojection~\cite{liu2013limited}. 

\subsubsection{Sparse View}
\begin{figure}[!htb]
    \centering
    \includegraphics[scale=0.37]{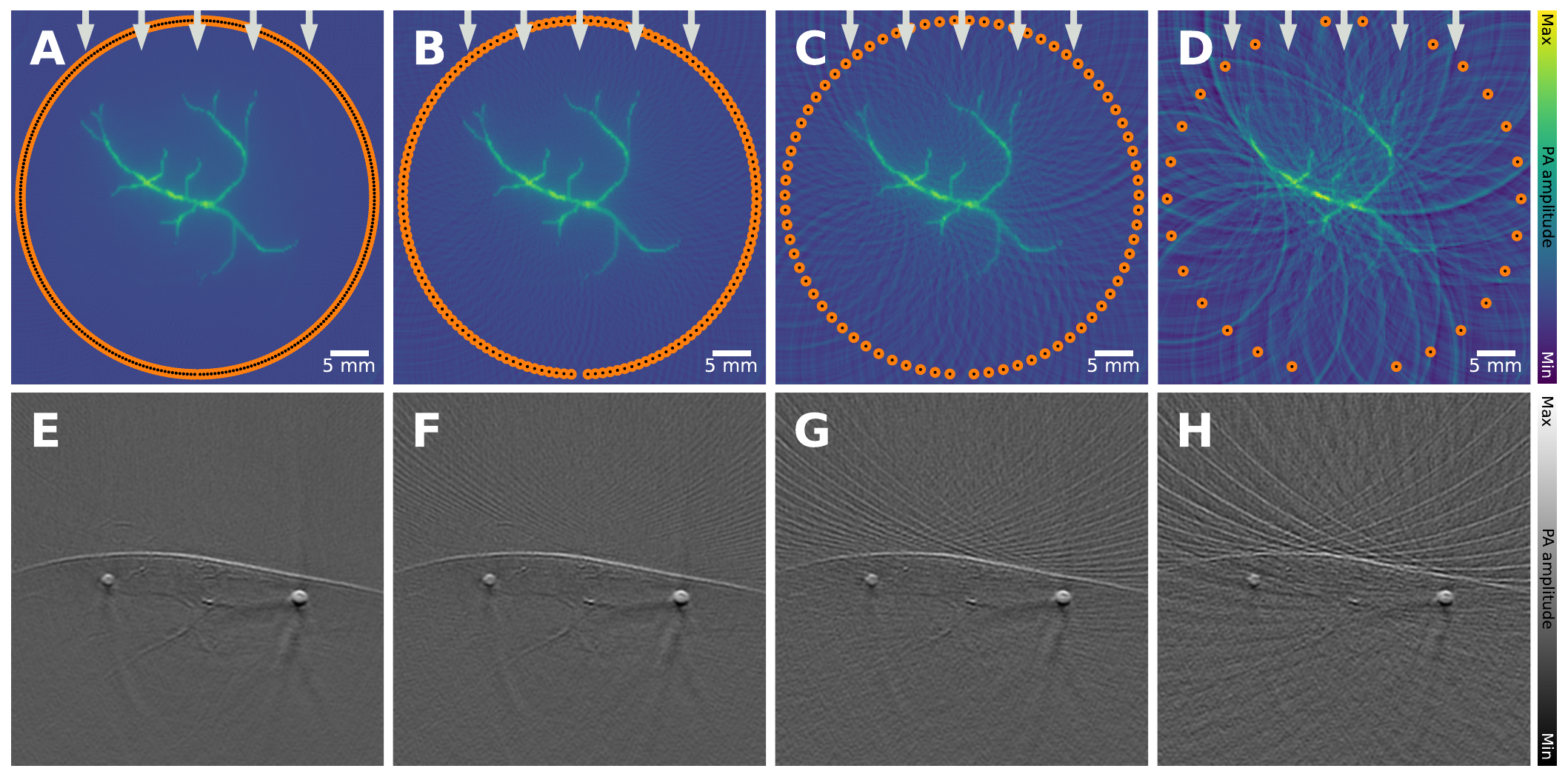}
    \caption{\textbf{Sparse view artifact: leads to clutter and signal loss.}  A vascular network is imaged under identical settings, except for the detector density changes. As the number of detectors becomes smaller (\textbf{A}: 298, \textbf{B}: 148, \textbf{C}: 74, \textbf{D}: 28) the image quality increasingly drops. The increase in the characteristic streaking artifact is visible both in the simulations (\textbf{A--D}) as well as \textit{in vivo} images (\textbf{E--H}). The \textit{in vivo} images show a human forearm imaged with a semicircle array (radius: 40 mm and 180\degree{} opening angle), \textbf{E}: 256, \textbf{F}: 128, \textbf{G}: 64 and \textbf{H}: 32 elements. \textbf{E--H} have been reprinted from Ozdemir et al., 2022~\cite{ozdemir2022oadat}, which is available open-access under a CC BY 4.0 license.}
    \label{fig:hardware:sparse}
\end{figure}

A sparse arrangement of detection elements can be necessary for PAI hardware systems, for example, due to budget considerations~\cite{guan2020limited}, when attempting to maximize the imaging speed~\cite{ozdemir2022oadat}, or by attempting to minimize the hardware or computational complexity~\cite{huang2023unveiling}. Spatial under-sampling also gives rise to grating lobes. There are additional beams created by an array transducer and result from constructive interference caused by periodic spacing of the elements when the element spacing is greater than half the wavelength of the ultrasound wave~\cite{paul1997side}. Such spatial under-sampling leads to characteristic streaking artifacts, which can be seen both in idealized simulations (Fig.~\ref{fig:hardware:sparse}\textbf{A--D}) but also on \textit{in vivo} images (Fig.~\ref{fig:hardware:sparse}\textbf{E--H})~\cite{ozdemir2022oadat}. In some situations, it can be possible to address the problem through view interpolation or via the use of a carefully chosen regularization term during iterative reconstruction, such as enforcing a total variation constraint~\cite{zhang2012total}.

% \begin{figure}[!htb]
%     \centering
%     \includegraphics[scale=0.37]{4_directivity.pdf}
%     \caption{Detector directionality artifact. A vascular network is imaged under identical settings, except the detector directionality changes. Assumption is that of point-like detectors, which is valid for the top row. For the two rows below, the detectors increase in length and become increasingly exclusively sensitive for waves along the imaging direction, decreasing image qualify.}
%     \label{fig:directivity}
% \end{figure}

\subsection{Response of the Detector Elements}
There is also a limit to the efficiency with which detection elements can measure the acoustic pressure in a medium. This efficiency depends on three main factors: the incident angle of the wave onto the detection element (referred to as the \textit{detector directivity}), the sampling rate of the detection element (referred to as the \textit{temporal sampling}), and the frequency components of the wave (referred to as the \textit{frequency response}). 

\subsubsection{Detector Directivity}
We often assume that detector elements are perfectly omnidirectional, but in reality, detectors are sensitive to the direction of the incoming sound waves. To achieve an omnidirectional response, infinitely small point pressure detectors would be required. Real detectors have a finite aperture size, however, and are most sensitive to sound waves arriving perpendicular to their surface and gradually less sensitive with oblique angles~\cite{cox2010effect}. When ignored in the image reconstruction, the finite aperture size leads to a significant blurring effect in the reconstructed images, where the degree of the blurring increases with the aperture size~\cite{warbal2022silico}. A high directivity (implemented by a large aperture size) may be beneficial to reduce the influence of out-of-plane contributions, and line detectors with extreme dimensions can even reduce the reconstruction to a Radon transform~\cite{paltauf2017piezoelectric}. However, the image quality can be significantly impacted by a high directivity, with an increased degree of blurring. In practice, we can have a large element size in the elevation direction to give a thin image plane, while having a small element size in-plane, retaining a high resolution~\cite{cox2010effect}. It should also be noted that detector directivity is frequency-dependent.

\subsubsection{Limited Temporal Sampling}
\begin{figure}[!htb]
    \centering
    \includegraphics[scale=0.37]{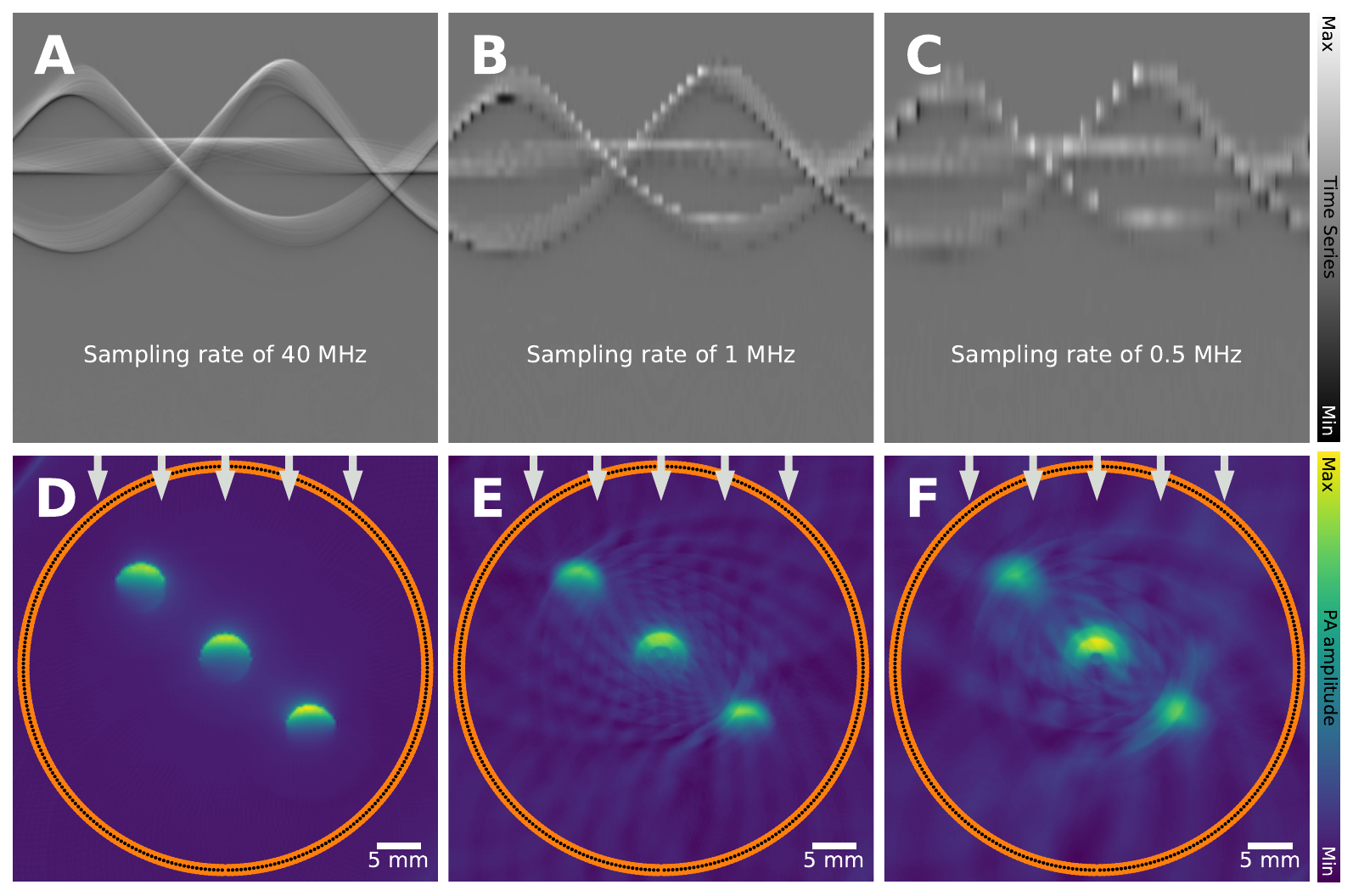}
    \caption{\textbf{Limited temporal sampling artifact: leads to blurring, clutter and signal loss.} 3 blood vessels are imaged under identical settings, except the temporal sampling rate changes. As the Nyquist sampling theorem is increasingly violated (\textbf{A}: 40MHz, \textbf{B}: 1MHz, \textbf{C}: 0.5 MHz), artifacts occur.}
    \label{fig:hardware:sampling_rate}
\end{figure}

The sampling rate of an imaging system inherently limits the maximum detectable frequency as apparent from the Nyquist sampling theorem~\cite{nyquist1928certain}. The theorem states that one has to sample at twice the frequency that should still be detectable. All frequency components above this threshold can lead to undersampling (or aliasing) artifacts. Most conventional ultrasound data acquisition units use a sampling rate of (at least) 40 MHz, which means that the maximum detectable frequency in the signal is limited to 20 MHz. In cases where the sampling rate is below the major frequency components of the acoustic waves, significant aliasing artifacts and blurring can be expected, which can be seen in Fig.~\ref{fig:hardware:sampling_rate}. In some situations, it can be possible to mitigate these artifacts, for example through the application of a simple frequency-domain filter~\cite{gamelin2011fast}, or through compressive sensing techniques~\cite{arridge2016accelerated}.

\subsubsection{Limited Frequency Response}
\begin{figure}[!htb]
    \centering
    \includegraphics[scale=0.37]{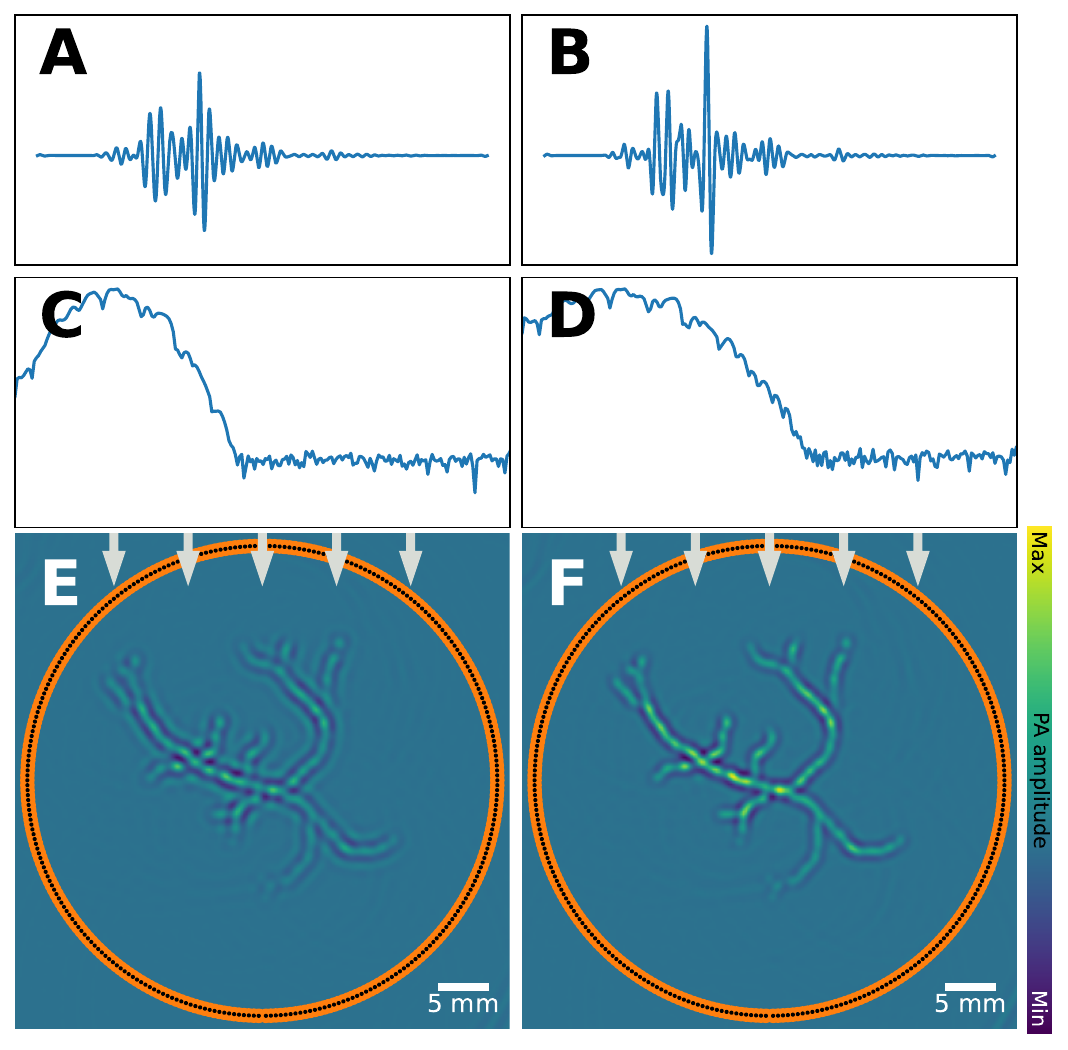}
    \caption{\textbf{Limited frequency response artifact: leads to blurring, clutter and signal loss.} A vascular network is imaged under identical settings, except the detector bandwidth changes. Transducers have a limited bandwidth, here modeled as a frequency domain Gaussian filter with a certain bandwidth and a center frequency. \textbf{A}: 1 MHz, 0.5 MHz, \textbf{B}: 1MHz, 0.8MHz: center frequency and bandwidth respectively. We show the time series data trace for one detection element (\textbf{A}\&\textbf{B}), the frequency content of the measured signals (\textbf{C}\&\textbf{D}), and the reconstructed vascular network (\textbf{E}\&\textbf{F}). One can clearly see differences in the signal amplitudes and the sharpness of the reconstructed vasculature.}
    \label{fig:hardware:impulseresponse}
\end{figure}

The temporal sampling discussed above introduces a natural upper bound of the detectable frequency components of the acoustic waves. Furthermore, different types of detectors have varying sensitivity to the frequency components of the incoming waves. The relationship of sensitivity and frequency is referred to as the \textit{impulse response} of the system. It describes how the imaging devices react to a sharp rise and fall in sound pressure (an ideal Dirac impulse) as a function of sensitivity over wavelength. In the case of piezoelectric ultrasound transducers, these are typically characterized using a center frequency and a bandwidth. Depending on the impulse response of the imaging system, measurements of the same sound waves can drastically vary, which can be seen in Fig.~\ref{fig:hardware:impulseresponse}.

Capturing high-frequency components of the ultrasound wave is important to reconstruct a sharp image, but noise also manifests significantly in the higher frequencies. To mitigate this, low-pass filters are often applied prior to image reconstruction, which can mitigate noise but have to be traded off with a potential blurring of the image. Conversely, if the impulse response function limits the capture of lower frequencies, information can be lost from smoothly varying or piecewise-constant regions. Several mitigation strategies have been proposed for the loss of low frequencies, including deconvolution with~\cite{rejesh2013deconvolution,van2016comparison} or without~\cite{wang2004photoacoustic} known impulse response, or by using deep learning-based data priors for correction~\cite{gutta2017deep,munjal2024deep}.

\subsection{Measurement Noise}
\begin{figure}[!htb]
    \centering
    \includegraphics[scale=0.37]{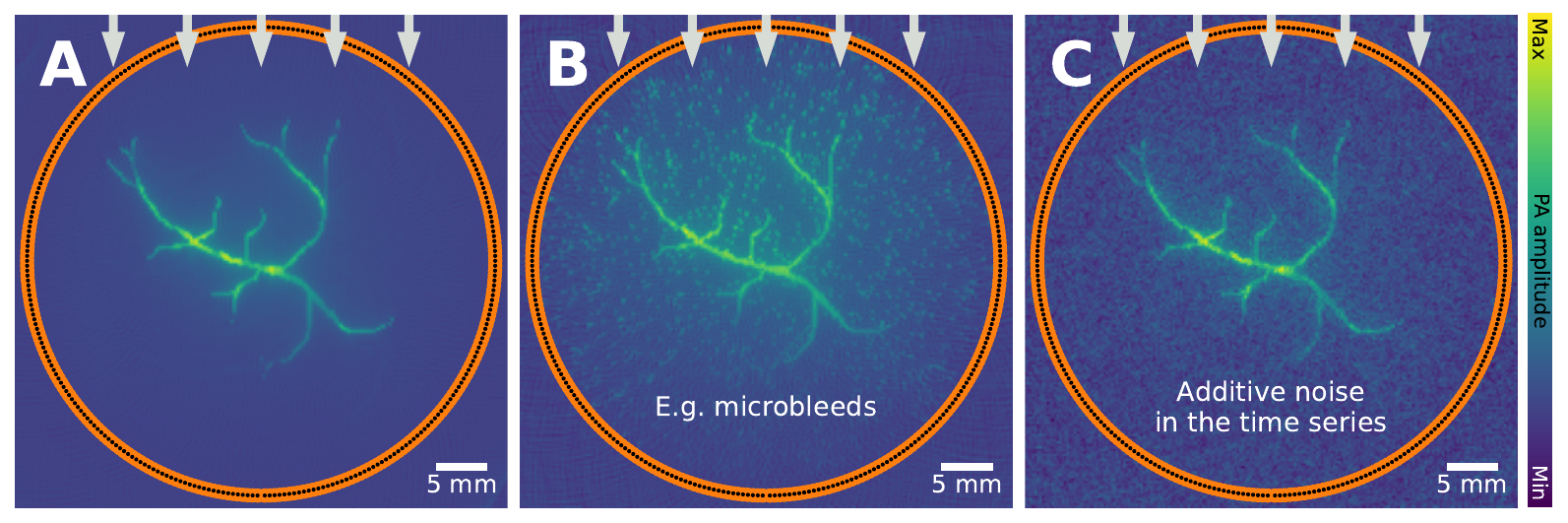}
    \caption{\textbf{Measurement noise artifact: leads to signal loss at low SNR, and clutter.} A vascular network is imaged under identical settings, except noise is introduced in \textbf{B} and \textbf{C}. \textbf{B}: small clusters of blood are introduced (e.g.\ due to microbleeds), \textbf{C}: noise (zero-mean additive Gaussian noise, with a standard deviation of 0.05) is added to the time series. Both result in a similar reconstructed image, which can complicate image analysis.}
    \label{fig:hardware:noise}
\end{figure}

Stochastic changes to the measurements can arise from several sources, including thermal and electrical effects~\cite{winkler2013noise}. The amount of noise in an image is often quantified using specific measures, such as the signal-to-noise ratio (SNR), the contrast-to-noise ratio (CNR)~\cite{welvaert2013definition}, or the generalised contrast-to-noise ratio (gCNR)~\cite{kempski2020application}. For all of these measures, the relative signal differences in two regions of interest (one corresponding to noise, and one to the target signal) are analyzed. Common mitigation strategies include the use of data filtering techniques~\cite{hu2023adaptive} or by using dimensionality reduction techniques to separate meaningful signal from noise~\cite{tzoumas2014spatiospectral,holan2008automated}. However, as with other artifacts introduced in this review, measurement noise may not always be clearly differentiable from real signals, see Fig.~\ref{fig:hardware:noise}. For example, if the distribution of signals is high-frequency, a reconstructed image of high Gaussian noise might look qualitatively comparable. Additionally, detectors have a noise floor, referred to as the \textit{noise-equivalent pressure}~\cite{winkler2013noise}, which quantifies the minimum acoustic pressure fluctuations that are measurable.

\clearpage % force Latex to drop any held images here, so we can keep the sections separate while editing.

\section{Artifact Source: Patient}
Properties of human subjects, such as the presence of highly optically absorbing structures like melanin in the epidermis, or structures with very different acoustic properties, such as bone and soft tissue underlie many of the artifacts already introduced. Additionally, however, human imaging poses unique challenges that do not fall under of the previous categories. 

\subsection{Patient Movement}
Patient movement poses a substantial challenge for accurate and high-quality clinical PAI. Motion artifacts depend closely on the imaging speed, so systems with low laser pulse repetition rates or slow ultrasound acquisition rates are particularly affected by this. Furthermore, certain anatomical sites are more susceptible to motion artifacts because of breathing and heartbeat-induced motion, for example, near the lungs in breast imaging, or melanoma imaging of the torso or back~\cite{Lin2018}. Mitigation strategies are therefore crucial to enable reliable clinical imaging of these organs.

\begin{figure}[!htb]
    \centering
    \includegraphics[scale=0.37]{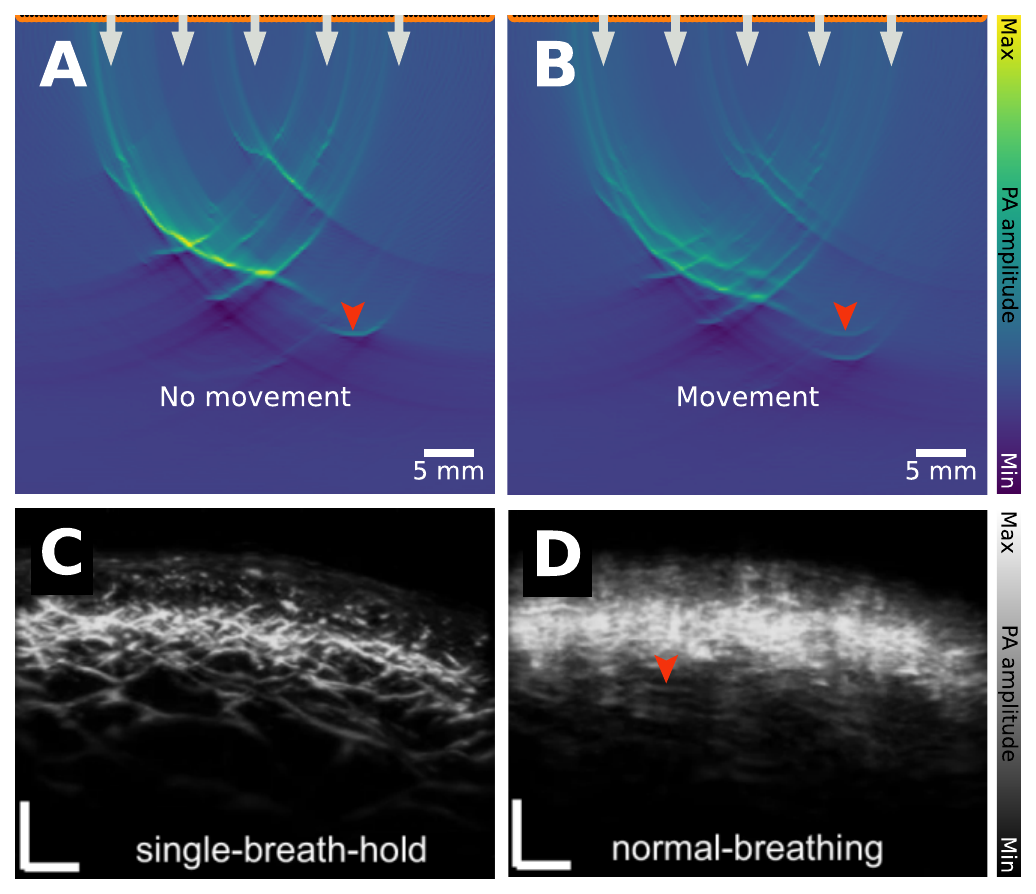}
    \caption{\textbf{Patient movement artifact: leads to clutter and blurring.} Measurements are often repeated, e.g.\ to average and reduce random noise. The assumption is that the patient is stable and does not move. \textbf{A}: the assumption holds, where 3 (identical) time series of the vascular network are recorded and averaged. \textbf{B}: there is movement between frames, and the 3 time series differ, resulting in decreased image quality.
    This can also be seen in \textbf{in vivo} data (\textbf{C}\&\textbf{D}), where the skin on the back of a healthy volunteer is imaged. \textbf{C}: acquired while the breath was held for 15s (single-breath-hold), and \textbf{D}: acquired during normal breathing.
    The \textit{in vivo} images have been reprinted from He et al., 2022~\cite{he2022fast}, which is available open-access under a CC BY 4.0 license.}
    \label{fig:patient-movement}
\end{figure}

The effects of patient motion on PAI depends on the exact image acquisition procedure, such as the resolution, frame rate and post-processing procedure. In a system with a fixed array of transducers, such as typical tomographic clinical systems, each frame is acquired in less than $50~\mathrm{\mu s}$, as determined by the sound speed and field of view. Therefore, for any substantial artifact to be induced over the course of a single frame, motion greater than the resolution limit would have to occur on that time frame. For a typical resolution of $100~\mathrm{\mu s}$, this would require motion faster than around $2~\mathrm{m/s}$, which is much faster than even blood flow in arteries ($\sim 0.3 \mathrm{~m/s}$~\cite{leeGeneralPrinciplesCarotid2013}), so no significant artifact is expected in a single-shot frame. Multi-wavelength imaging, or averaging across multiple frames, however, occurs on a much slower time scale ($\sim 10\mathrm{~Hz}$ for tuneable laser systems), where motion between wavelengths is certainly possible, see Fig.~\ref{fig:patient-movement}. Motion due to probe movement by the operator or breathing motion could lead to features like blood vessels appearing in different locations depending on wavelength and split or blurred in the case of frame averaging. If spectral unmixing is then applied to resolve the contributions of different biomolecules to the photoacoustic spectrum, the motion of the patient means that the spatial location of a given pixel will have changed during the scan, rendering spectral unmixing results meaningless.

Several strategies exist to mitigate the influence of motion on PA data. One straightforward approach is breath-hold imaging, which minimizes motion-related distortions in thoracic and abdominal imaging by instructing the subject to hold their breath during data acquisition~\cite{Lin2018} (Fig.~\ref{fig:patient-movement}\textbf{C}). However, this is not always feasible, especially in clinical settings involving long imaging times. Computational methods offer alternative solutions, such as the approach demonstrated by Schwarz et al.~\cite{SCHWARZ2016375} in Raster-Scan Optoacoustic Mesoscopy (RSOM, iThera Medical GmbH, Germany), where motion correction is achieved by identifying the position of the melanin layer in the skin and using it to correct the wavefront distortions. Another method involves selective frame averaging~\cite{avihai2019self}, where adjacent frames are compared based on an image quality metric, and averaging is only applied when images meet a predefined similarity threshold. This technique helps reduce motion blur without introducing excessive loss of spatial resolution. Additionally, optical flow-based correction has been proposed to track motion across frames by computing a deformation map. However, this method faces challenges in multi-wavelength imaging, where spectral variations alter image features. A possible solution is to compute the optical flow map on a co-registered ultrasound image, then apply the derived deformation map to the PA scan, thus improving the robustness of motion correction across different wavelengths~\cite{kirchner2019open}.

\subsection{Patient Preparation}
\begin{figure}[!htb]
    \centering
    \includegraphics[scale=0.37]{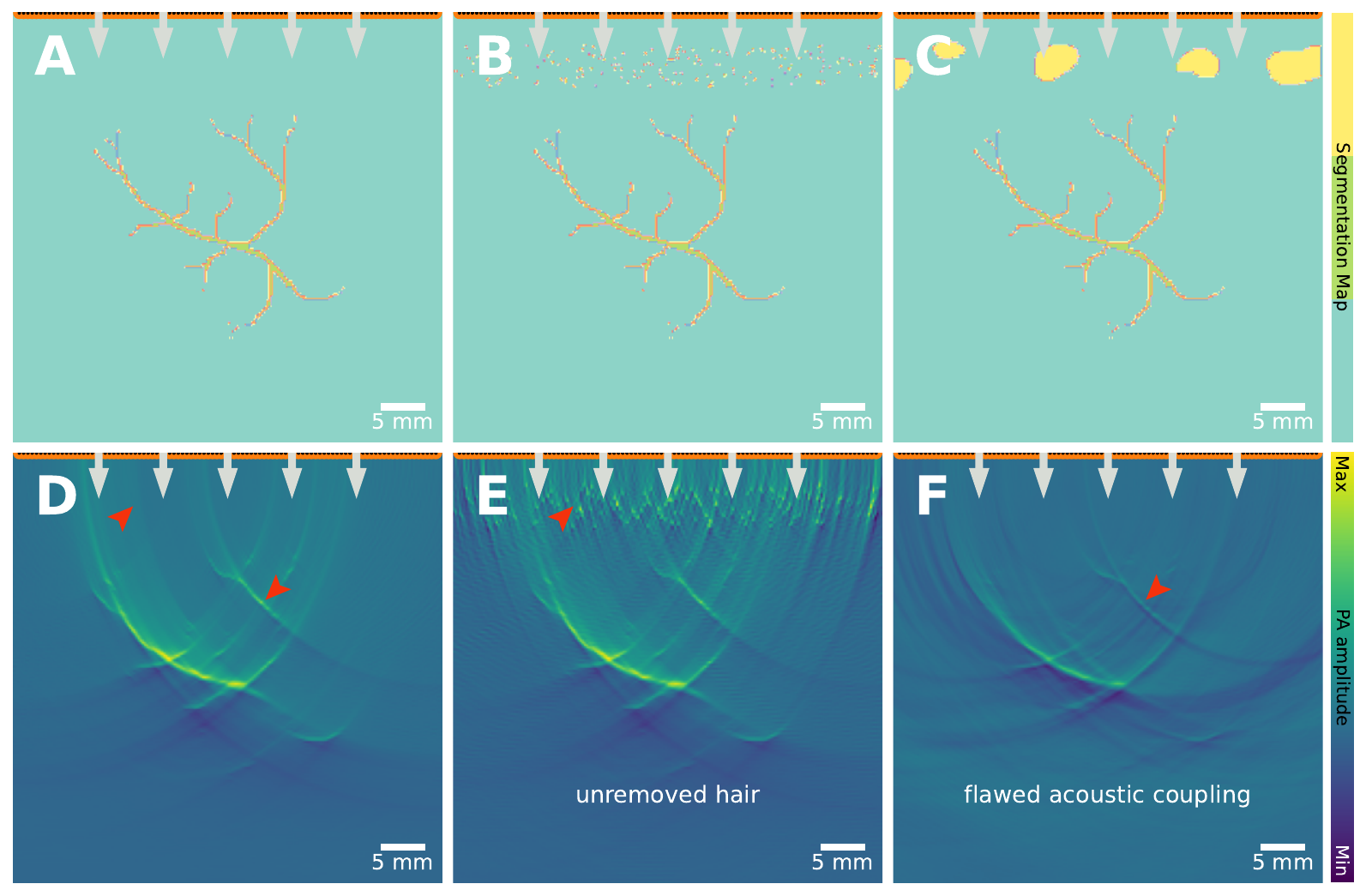}
    \caption{\textbf{Patient preparation artifact: leads to clutter.} A vascular network is imaged under identical settings, except for changes in quality of patient preparation. \textbf{B}: shows hairs that should have been removed before measurement, resulting in strong absorption and decreased image qualify when compared to the ideal case in \textbf{A}. \textbf{C}: shows air bubbles between transducer and object, which strongly distort the time series and the resultant image.}
    \label{fig:patient-prep}
\end{figure}

Lack of correct preparation of the patient can result in artifacts, as shown in Fig.~\ref{fig:patient-prep}. For example, hair on the surface of the skin can strongly absorb light, particularly in darker hair types~\cite{fordStructuralFunctionalAnalysis2016}. Due to the limited-view artifacts induced in practical clinical PAI systems, absorption in the skin surface will lead to artifacts below the skin surface (Fig.~\ref{fig:patient-prep}\textbf{B}\&\textbf{E}). This could qualitatively and quantitatively affect the image below the skin surface, by introducing extraneous signal, or reducing the contrast between the background, non-absorbing tissue and the target of interest. To mitigate this source of artifacts, a suitable hair-removal protocol should be followed, by shaving, waxing, or using depilatory cream on the imaging site. 

Air bubbles (which can be present in e.g.\ coupling gel) lead to large spatial differences in the sound speed and density. The differences in acoustic properties between biological tissue and the air bubbles lead to acoustic reflections, violating the assumption of straight-line unimpeded acoustic propagation made in the image reconstruction. Violation of these image reconstruction assumptions means that the image is blurred, the signal intensity is reduced, and sometimes regions of very low or negative signal intensity can be seen (Fig.~\ref{fig:patient-prep}\textbf{C}\&\textbf{F}). Similar artifacts may be observed when imaging gas-filled structures in the body, such as the intestines, as highlighted in images of phantom materials without proper degassing of the material~\cite{hacker2023fabrication}.

\section{Advanced Mitigation Strategies}
\label{sec:adv_mitigation}
A plethora of PA image reconstruction algorithms have been proposed to recover initial pressure distributions from PA time series measurements. We do not attempt a comprehensive review here of a substantial body of work, but note in passing that the various approaches include one-step filtered backprojection-type methods~\cite{xu2005universal,kunyansky2007explicit}, 
beamforming~\cite{perrot2021so,matrone2014delay,jaeger2007fourier}, time reversal~\cite{finch2004determining,burgholzer2007exact,hristova2008reconstruction},
iterative reconstruction schemes
~\cite{agranovsky2007uniqueness,stefanov2009thermoacoustic,qian2011efficient},
optimisation approaches using numerical models of acoustic propagation\cite{wang2012investigation,arridge2016accelerated,cox2017modeling}, and more recently deep learning-based reconstruction schemes~\cite{hauptmann2018model,waibel2018reconstruction,kim2020deep}. 
The important point is that all these reconstruction methods incorporate a model approximating the physics of the acoustic propagation and detection. While this model may not always be explicitly stated, it is inherently present.
Inaccuracies in the physics model are one of the two fundamental reasons why artifacts occur in images; the other being insufficient data. Corresponding to these two fundamental causes of the artifacts are two corresponding classes of approaches to mitigating artifact generation, which we describe in the next sections, as well as deep learning and deconvolution-based approaches.

\subsection{Improving the model}
If the artifacts are due to inaccurate approximations in the physical model, improving the physical model may remove them. For example, the approximation that the sound speed has no spatial variation is not always true and the resulting artifacts can be ameliorated by incorporating a more accurate representation of the sound speed in the model used in the reconstruction. There are two challenges to this: knowing what the correct physical model should be (in this example, knowing the correct sound speed map), and incorporating it into the reconstruction algorithm. 
Two-sound speed models have been used to reduce blurring when the coupling medium has a different sound speed from the tissue and the two regions and their average sound speeds are known a priori, or the regions can be segmented from a preliminary image and literature values used for the sound speeds~\cite{dean2017accounting,yue2022double}. However, in more general cases, the sound speed distribution must be estimated using ultrasound tomography~\cite{jose2012speed,xia2013enhancement,wurzinger2016combined,mervcep2019transmission,dantuma2023fully}. Once a model of the sound speed distribution has been obtained, by whatever means, the challenge is to incorporate it into the PA reconstruction. In the two-sound speed case, backprojection methods can be readily adapted. Models that can account for more general sound speed maps include series-based approaches
\cite{agranovsky2007uniqueness,stefanov2009thermoacoustic,qian2011efficient} or iterative image reconstruction approaches based on the minimization of a functional (an objective, cost or loss function) that measures the closeness of fit of a numerical model to the data~\cite{wang2012investigation,arridge2016accelerated,belhachmi2016direct, tick2019modelling}. In this latter case, the main challenges become keeping the computational burden manageable and accelerating the convergence of the minimization. For small data sets and image sizes, it is possible to form the required operators explicitly in matrix form and exploit recent advances in linear algebra but these tasks become non-trivial for large 3D images and data sets.
Sound speed estimation from the PA images or PA data itself has been proposed, but, while interesting, it is still under investigation~\cite{zhang2006reconstruction,cui2021adaptive,matthews2017joint,poudel2020joint,jeong2025revisiting}.

\subsection{Supplementing the data}
If the artifacts are due to insufficient data, the ideal solution is to record more data, but that is not always possible. However, sometimes the artifacts can be removed by restricting the class of possible images during the image reconstruction stage by incorporating additional information, e.g.\ prior knowledge of the object, such as that it must be non-negative, or piecewise constant, or take a certain form. 
Some prior information can be used in a post-processing step, e.g.\ by using a vessel filter~\cite{frangi1998multiscale} when the image is known to contain vessels, or applying a non-negativity condition when the physical principles dictate that it must be the case. Priors may also be obtained from adjunct imaging modalities, often ultrasound imaging~
\cite{mandal2019multimodal,yang2020soft,zhao2022ultrasound}.
However, functional-minimisation image reconstruction approaches, mentioned above, offer a more systematic approach as additional terms, \textit{penalty} terms, can be added to the functional to promote images of a certain type. The total variation approach, which promotes piecewise constant solutions, is commonly implemented in this way. A related approach will ensure an image satisfies a prior by iterating between updating an image estimate based on the data and projecting that image into the space of allowed solutions using a proximal operator~\cite{rudin1992nonlinear,yao2011enhancing,zhang2012total,wang2012investigation, parikh2014proximal, arridge2016accelerated, boink2017reconstruction}.

\subsection{Deep learning}
Because model-based functional minimization methods for image reconstruction can be computationally intensive, deep learning networks, which are slow to train but fast to run, have been proposed as a means of speeding up reconstruction times. 
Furthermore, learned components can be used at every step of the image reconstruction pipeline: to preprocess the data, e.g.\ to remove noise, to generate the initial guess of an iterative reconstruction, to act as a model of the physics, to learn a prior to compute an image update, to post-process the image e.g.\ to remove artifacts, and even to replace the whole image reconstruction pipeline with a learned end-to-end data-to-image reconstruction. The use of deep learning in medical image reconstruction in general, as well as in photoacoustics, is a fast-changing field. We will therefore not include references here, other than to point the interested reader to some review articles~\cite{hauptmann2020deep,grohl2021deep,deng2021deep,yang2021review,rajendran2022photoacoustic,yang2023recent,wei2024deep}.
As this is a paper on artifacts, it must be noted that any learned method is only as good as the data used to train it, and that over-reliance on the learned distribution in preference to the measured data can result in \textit{hallucinations}: artifacts generated by the learned component and not arising from the data.

\subsection{Deconvolution}
Deconvolution is widely used in data space as a method for correcting for the frequency response of acoustic response of the detector elements~\cite{Yi_Wang_2004}, as well as in image space to correct for the point-spread function (PSF) of the PAI system~\cite{9420701}. Deconvolution is essentially a re-weighting of the frequency components of a signal (or spatial frequency components of an image) to account for the non-ideal response of the detector or imaging system. This can be beneficial, as attenuated frequency components can be restored to their correct amplitude. However, care must be taken. First, when a frequency component has been attenuated to the extent that is it indistinguishable from the noise, or when it has simply not been measured because of the limitations of the system, naive deconvolution will amplify the noise, corrupting the image. Regularization of the deconvolution is therefore necessary. Second, when deconvolving an image PSF from an image, it is often assumed that the PSF is spatially invariant. This allows the deconvolution to be applied efficiently using Fast Fourier Transforms. However, this assumption of spatial invariance will not always hold, for instance, limited-view artifacts are not usually spatially invariant. Note that image reconstruction schemes in which the model of the detection system (e.g.\ the frequency- and directional-dependence of the detector elements) is included in the model of the physics used in the reconstruction, will effectively, implicitly, deconvolve the PSF from the image during the image reconstruction process.

\section{Discussion}
As photoacoustic imaging makes its way from the laboratory to the clinic, the presence of artifacts could adversely affect patient care if not identified or corrected. Here, we have presented an overview of the mechanisms responsible for artifact generation and illustrated them, providing a taxonomy and example images. This can serve as a guide for clinical users, but also developers of novel PA systems.

Artifacts in PAI have substantial implications for its clinical use, where we hope that our work can be of use for furthering the understanding and mitigation options for commonly encountered artifacts. For example, one of the FDA-approved PAI systems calculates oxygenation maps, which can be used as a biomarker in tumor diagnosis
\cite{kratkiewicz2022ultrasound}. Various light-tissue interaction artifacts can distort this oxygenation map, such as spectral colouring and out-of-plane absorption. The distortion of the oxygenation map can, in turn, lead to a misdiagnosis of the tumor if clinical users are not familiar with this inherent limitation. Moreover, as the sound speed varies substantially in human tissue, from 1450 m/s for fat to 4080 m/s for bone~\cite{doi:10.1148/rg.294085199}, and also between pathology~\cite{bamber1981acoustic}, sound speed artifacts are likely to occur in PAI. These artifacts will hamper the investigation of the size, shape, and characteristics of structures, which could mean incorrect risk stratification~\cite{10.1117/1.JBO.29.S1.S11515}. In contrast, artifacts may also carry diagnostic information. For example, in ultrasound, the so-called comet-tail artifact can be used for finding gallstones~\cite{https://doi.org/10.7863/jum.1982.1.1.1}, and in CT the blooming artifacts can highlight calcifications~\cite{Park2024}. Our overview of artifacts in PAI can therefore help with preventing confounding during the diagnostic process, but also assist in the diagnostic process.

By listing the limitations of current PAI systems and demonstrating the origins of the resulting artifacts, we hope that this work will assist those working to improve PAI in the future. Previously, in-depth knowledge of the out-of-plane absorption artifact allowed for the formulation of a transducer displacement method for reducing their influence~\cite{Nguyen:19}. As mentioned, sound speed artifacts have also been ameliorated by integrating ultrasound tomography into photoacoustic systems, such that the measured sound speed maps can be used during the reconstruction~\cite{10.1117/12.3043362}. Recently, deep learning methods have also been introduced in combination with PAI, either as post-processing correction algorithms~\cite{waibel2018reconstruction} or by directly learning the reconstruction~\cite{guan2020limited}, but (artifact) hallucinations may hamper their usefulness. Physics-driven deep learning, where one embeds knowledge of physical laws, has shown high-quality results in PAI~\cite{SHEN2024}.

While we have discussed many artifacts, the scope of this paper is limited to artifacts that originate outside of the reconstruction step. There are different reconstruction algorithms for PAI, each with their own advantages and disadvantages~\cite{10.3788/PI.2024.R06}. While a reconstruction algorithm may dampen the impact of certain artifacts, it can also increase the impact of others or even introduce new artifacts. These reconstruction artifacts are often characteristic of the chosen algorithm, as they handle data limitations differently. For example, some algorithms may result in negative values in the reconstructed image~\cite{Tian_2021}, backprojection has shown to result in more pronounced streak artifacts~\cite{Cai:19}, and deep learning techniques could induce artifacts by hallucination. In this work, we did not systematically investigate the influence of the choice of reconstruction algorithm on the reconstructed image quality, though that would undoubtedly be of great interest to the PAI community.

In addition to differences in reconstruction algorithms, there also exist many \textit{ad-hoc} pre- and postprocessing schemes that are widely used. Such computational steps usually significantly change the image content and alter how the data is perceived by the user. For example, the application of vesselness filters (such as the Frangi vesselness filter~\cite{frangi1998multiscale}) is quite common in PAI~\cite{oruganti2013vessel}, however, such an approach can provide inconsistent and possibly misleading results when applied to PA images obtained under realistic conditions~\cite{longo2020assessment}. Other approaches include the application of a variety of false-color scales, which can non-linearly affect the contrast of image features. These are just two common examples, but all of these methods come hand-in-hand with their own characteristic shortcomings that a user of the technology needs to keep in mind when, for example, basing clinical decisions on the images.

\section {Summary}
An idealized photoacoustic system would provide an image that corresponds spatially and spectrally to the morphological and molecular properties of the tissue. However, modeling assumptions must often be made in image reconstruction and post-processing to ensure computational tractability, leading to artifacts where the assumptions break down. Furthermore, clinical requirements, physics constraints and practical limitations can mean that incomplete data is provided. As the image reconstruction algorithms place several requirements on the amount of data collected, these incomplete datasets will lead to artifacts. Artifacts can be mitigated with certain strategies, such as other reconstruction models, more complete datasets, and including prior knowledge in the reconstruction, but currently, artifacts can not be prevented entirely. Artifacts have the potential to adversely affect results by introducing false features or obscuring true features, significantly hampering the clinical translation of PAI. We believe that an in-depth awareness and knowledge of the origins of PA artifacts is a critical step to catalyze future innovations for the development of mitigation strategies.

\section*{Acknowledgments}
MTR and SM acknowledge the funding from KWF, TKI-Life Sciences and Health and Seno Medical Instruments in project THYNAS+. SM acknowledges funding from EFRO-Oost project 00103 Elastografie voor snellere herkenning van borstkanker in 3D fotoakoestieche mammografie. The work of JG was supported by the Deutsche Forschungsgemeinschaft (DFG, German Research Foundation) under projects GR 5824/1 and GR 5824/2. SEB and TRE were supported by Cancer Research UK under grant number C9545/A29580 and the UKRI Engineering and Physical Sciences Research Council under grant numbers EP/X037770/1, EP/R003599/1 and EP/V027061/1. BTC acknowledges support from the Engineering and Physical Sciences Research Council, UK (EPSRC), grants EP/W029324/1, EP/T014369/1.

\section*{Author contributions}
All authors contributed to the conception of the project idea. The initial draft of the manuscript was written by MTR, JMG, TRE, SM, and BTC. MTR developed the code used for data simulation and designing the figures. Following the initial draft, all authors reviewed the manuscript, providing feedback and input to shape the final version.

\bibliographystyle{elsarticle-num} 
\bibliography{literature}

\newpage
\section*{Supplementary information}
% If your article requires supplementary information, please include these files for peer-review. Please note that supplementary information will not be edited.
\begin{table}[!htb]
\begin{tabular}{|ll|}
\hline
\multicolumn{2}{|l|}{\textbf{Modeling assumptions: light and sound physics}} \\ \hline
\multicolumn{1}{|l|}{\textbf{L1}} & Fluence and radiant exposure are not spatially varying. \\ \hline
\multicolumn{1}{|l|}{\textbf{L2}} & Fluence does not depend on absorption. \\ \hline
\multicolumn{1}{|l|}{\textbf{L3}} & Fluence is wavelength-independent. \\ \hline
\multicolumn{1}{|l|}{\textbf{L4}} & All the absorbed energy is translated into an acoustic wave. \\ \hline
\multicolumn{1}{|l|}{\textbf{S1}} & Sound waves do not decay with depth. \\ \hline
\multicolumn{1}{|l|}{\textbf{S2}} & The sound speed is uniform in tissue. \\ \hline
\multicolumn{1}{|l|}{\textbf{S3}} & Sound waves are not scattered in tissue. \\ \hline
\multicolumn{1}{|l|}{\textbf{S4}} & Sound waves behave independently of their frequency. \\ \hline
\multicolumn{2}{|l|}{\textbf{Data assumptions: excitation and detection hardware}} \\ \hline
\multicolumn{1}{|l|}{\textbf{H1}} & There is sufficient data for exact image reconstruction (spatio-temporal sampling, frequency bandwidth). \\ \hline
\multicolumn{1}{|l|}{\textbf{H2}} & Laser power is constant for each pulse. \\ \hline
\multicolumn{1}{|l|}{\textbf{H3}} & The input wavelength is known perfectly. \\ \hline
\multicolumn{1}{|l|}{\textbf{H4}} & Radiant exposure is uniform. \\ \hline
\multicolumn{1}{|l|}{\textbf{H5}} & Measurements are free of noise. \\ \hline
\multicolumn{1}{|l|}{\textbf{H6}} & Light pulse-length is sufficiently short. \\ \hline
\multicolumn{1}{|l|}{\textbf{H7}} & Detectors are perfectly directional. \\ \hline
\multicolumn{1}{|l|}{\textbf{H8}} & No signals are measured that have an origin out of the imaging field of view. \\ \hline
\multicolumn{2}{|l|}{\textbf{Experimental assumptions}} \\ \hline
\multicolumn{1}{|l|}{\textbf{E1}} & Acoustic coupling between device and the subject is perfect. \\ \hline
\multicolumn{1}{|l|}{\textbf{E2}} & Subject motion does not compromise image quality. \\ \hline
\multicolumn{1}{|l|}{\textbf{E3}} & Tissue properties and detector sensitivity are not affected by temperature and remain constant. \\ \hline
\multicolumn{1}{|l|}{\textbf{E4}} & Subject preparation or medication does not confound the target imaging biomarkers. \\ \hline
\multicolumn{1}{|l|}{\textbf{E5}} & There exist no patient-specific confounders (such as skin tone, BMI, sex, age). \\ \hline
\end{tabular}
\caption{Assumptions that are made during the PAI process}
\label{tab:assumptions}
\end{table}

\begin{landscape}
\begin{table}
\begin{tabular}{|l|l|l|cccccc|}
\hline
\multicolumn{1}{|c|}{\multirow{2}{*}{\textbf{Source}}} & \multicolumn{1}{c|}{\multirow{2}{*}{\textbf{Cause}}} & \multicolumn{1}{c|}{\multirow{2}{*}{\textbf{\begin{tabular}[c]{@{}c@{}}Violated \\ Assumptions\end{tabular}}}} & \multicolumn{6}{c|}{\textbf{Artifact Effect}} \\ \cline{4-9} 
\multicolumn{1}{|c|}{} & \multicolumn{1}{c|}{} & \multicolumn{1}{c|}{} & \multicolumn{1}{c|}{\textbf{Dislocation}} & \multicolumn{1}{c|}{\textbf{Blurring}} & \multicolumn{1}{c|}{\textbf{Clutter}} & \multicolumn{1}{c|}{\textbf{Signal loss}} & \multicolumn{1}{c|}{\textbf{Signal Change}} & \textbf{Splitting} \\ \hline
\multirow{2}{*}{Patient} & \textbf{\begin{tabular}[c]{@{}l@{}}Patient\\ Movement\end{tabular}} & E2 & \multicolumn{1}{c|}{X} & \multicolumn{1}{c|}{X} & \multicolumn{1}{c|}{} & \multicolumn{1}{c|}{} & \multicolumn{1}{c|}{} &  \\
 & \textbf{\begin{tabular}[c]{@{}l@{}}Patient\\ Preparation\end{tabular}} & E4 & \multicolumn{1}{c|}{} & \multicolumn{1}{c|}{} & \multicolumn{1}{c|}{X} & \multicolumn{1}{c|}{} & \multicolumn{1}{c|}{X} &  \\ \hline
\multirow{5}{*}{\begin{tabular}[c]{@{}l@{}}Light-Tissue\\ Interactions\end{tabular}} & \textbf{Fluence Decay} & L1, L2, H4 & \multicolumn{1}{c|}{} & \multicolumn{1}{c|}{} & \multicolumn{1}{c|}{} & \multicolumn{1}{c|}{With depth} & \multicolumn{1}{c|}{} &  \\
 & \textbf{Spectral Coloring} & \begin{tabular}[c]{@{}l@{}}L3, H3, H5, \\ E5, H2\end{tabular} & \multicolumn{1}{c|}{} & \multicolumn{1}{c|}{} & \multicolumn{1}{c|}{} & \multicolumn{1}{c|}{} & \multicolumn{1}{c|}{\begin{tabular}[c]{@{}c@{}}Between \\ wavelengths\end{tabular}} &  \\
 & \textbf{\begin{tabular}[c]{@{}l@{}}Out-of-plane\\ Absorption\end{tabular}} & H8 & \multicolumn{1}{c|}{} & \multicolumn{1}{c|}{} & \multicolumn{1}{c|}{X} & \multicolumn{1}{c|}{} & \multicolumn{1}{c|}{} &  \\
 & \textbf{Laser Power Variation} & H2, H3, E3 & \multicolumn{1}{c|}{} & \multicolumn{1}{c|}{} & \multicolumn{1}{c|}{} & \multicolumn{1}{c|}{} & \multicolumn{1}{c|}{X} &  \\
 & \textbf{Long Pulse Duration} & H6 & \multicolumn{1}{c|}{} & \multicolumn{1}{c|}{X} & \multicolumn{1}{c|}{} & \multicolumn{1}{c|}{} & \multicolumn{1}{c|}{} &  \\ \hline
The PA effect & \textbf{PA Efficiency} & L4, E3 & \multicolumn{1}{c|}{} & \multicolumn{1}{c|}{} & \multicolumn{1}{c|}{} & \multicolumn{1}{c|}{} & \multicolumn{1}{c|}{X} &  \\ \hline
\multirow{3}{*}{\begin{tabular}[c]{@{}l@{}}Sound-Tissue\\ Interactions\end{tabular}} & \textbf{Sound Speed} & S2, S3, E1 & \multicolumn{1}{c|}{X} & \multicolumn{1}{c|}{X} & \multicolumn{1}{c|}{} & \multicolumn{1}{c|}{} & \multicolumn{1}{c|}{} & X \\
 & \textbf{Acoustic Reflections} & S3, S4, E1 & \multicolumn{1}{c|}{} & \multicolumn{1}{c|}{} & \multicolumn{1}{c|}{X} & \multicolumn{1}{c|}{} & \multicolumn{1}{c|}{} &  \\
 & \textbf{Acoustic Attenuation} & S1, S3, S4 & \multicolumn{1}{c|}{} & \multicolumn{1}{c|}{} & \multicolumn{1}{c|}{} & \multicolumn{1}{c|}{With depth} & \multicolumn{1}{c|}{} &  \\ \hline
\multirow{6}{*}{Signal Detection} & \textbf{Limited View} & H1 & \multicolumn{1}{c|}{} & \multicolumn{1}{c|}{} & \multicolumn{1}{c|}{X} & \multicolumn{1}{c|}{X} & \multicolumn{1}{c|}{} &  \\
 & \textbf{Sparse View} & H1 & \multicolumn{1}{c|}{} & \multicolumn{1}{c|}{} & \multicolumn{1}{c|}{X} & \multicolumn{1}{c|}{X} & \multicolumn{1}{c|}{} &  \\
 & \textbf{Detector Directivity} & H7 & \multicolumn{1}{c|}{} & \multicolumn{1}{c|}{} & \multicolumn{1}{c|}{X} & \multicolumn{1}{c|}{X} & \multicolumn{1}{c|}{} &  \\
 & \textbf{\begin{tabular}[c]{@{}l@{}}Limited Temporal \\ Sampling\end{tabular}} & H1 & \multicolumn{1}{c|}{} & \multicolumn{1}{c|}{X} & \multicolumn{1}{c|}{X} & \multicolumn{1}{c|}{X} & \multicolumn{1}{c|}{} &  \\
 & \textbf{\begin{tabular}[c]{@{}l@{}}Limited Frequency\\ Response\end{tabular}} & S4, H1 & \multicolumn{1}{c|}{} & \multicolumn{1}{c|}{X} & \multicolumn{1}{c|}{X} & \multicolumn{1}{c|}{\begin{tabular}[c]{@{}c@{}}With spatial \\ frequency\end{tabular}} & \multicolumn{1}{c|}{} &  \\
 & \textbf{Measurement Noise} & H5, E3 & \multicolumn{1}{c|}{} & \multicolumn{1}{c|}{} & \multicolumn{1}{c|}{} & \multicolumn{1}{c|}{At low SNR} & \multicolumn{1}{c|}{X} &  \\ \hline
\end{tabular}
\caption{List of the sources, causes, and effects of artifacts typically encountered in PAI.}
\label{tab:artifact_list}
\end{table}
\end{landscape}

\end{document}